\documentclass[12pt,a4]{article}
\usepackage{amsmath, amssymb, slashed,graphicx,color,multicol}
\usepackage[toc,page]{appendix}
\usepackage[colorlinks=true,urlcolor=black,linkcolor=black,citecolor=black]{hyperref}
\usepackage[margin=1in]{geometry}
\usepackage[utf8]{inputenc}
\newtheorem{definition}{Definition}

 \numberwithin{lemmas}{section}

\newtheorem{thm}{Theorem}

\newcommand{\ubar} \underline

\newcommand{\lfn}{\mathrm{L}}
\newcommand{\one}{\hat{\mathbf{1}}}
\newcommand{\Gamme}{\mathsf{\Gamma}}
\newcommand{\vpsi}{{\boldsymbol{\psi}}}
\newcommand{\veta}{{\boldsymbol{\eta}}}
\newcommand{\vphi}{{\boldsymbol{\phi}}}

\newcommand{\s}{\mathcal{S}}
\newcommand{\g}{\mathcal{G}}
\newcommand{\h}{\hat{H}}
\newcommand{\hh}{\mathsf{\hat{H}}}
\newcommand{\uu}{\mathsf{u}}
\newcommand{\ww}{\mathsf{w}}
\newcommand{\sss}{\mathsf{s}}
\newcommand{\ee}{\mathsf{e}}
\newcommand{\yy}{\mathsf{y}}
\newcommand{\dprime}{{\prime\prime}}

\begin{document}

\title{Vacuum polarization in Uranium} 

\author{
A.N. Efremov
\\
Laboratoire de Chimie et Physique Quantiques (LCPQ), IRSAMC UMR5626\\
Université de Toulouse III and CNRS,\\
118 Route de Narbonne, F-31062 Toulouse, France\\
\\
\href{mailto:alexander@efremov.fr}{alexander@efremov.fr}
}

\maketitle

\begin{abstract}
We formulate the Hartree--Fock method using a functional integral approach. Then we consider a nonperturbative component of the vacuum polarization. For the Dirac--Coulomb operator the renormalization flow of the vacuum polarization is calculated numerically. For the Hartree--Fock operator the polarization is obtained by integrating an appropriately rescaled flow. The text includes an approximate calculation of the vacuum polarization in Uranium.
\end{abstract}


\section{An introduction}
One of the most important discoveries that laid a path to the contemporary QED is the Dirac equation. However this discovery introduced a notorious problem of the negative energy solutions which led Dirac to the creation of the hole theory. In order to preserve the probabilistic interpretation of the wave function and the particle-hole symmetry he gave a new definition of the  density matrix~\eqref{1911a}. A radical departure from this point of view proposed by R. Feynman. He was the first to realize that the negative energy solutions are necessary to describe the motion of the relativistic electrons~\cite{dic}. A more important difference with the quantum mechanics is the fact that the field operator in QED does not have any probabilistic interpretation. An excellent historical review can be found in "The Quantum Theory of Fields" by S.~Weinberg~\cite{sw}.

The subject of radiative corrections in atoms has a very rich history, see e.g.~\cite{bethe,wk,gravejat2013}. The reader can find a contemporary review of a variety of theoretical and experimental methods to calculate the transition energies of hydrogen-like, helium-like and lithium-like ions in  "QED tests with highly charged ions" by P.Indelicato~\cite{indelicato}. On the contrary in present work we focus on the vacuum polarization of the atom of uranium with all ninety-two electrons. To give a motivation for our definition of the partition function we select an electron and put it on the Fermi energy level. Of cause such an approximation breaks the original symmetry. To restore it we introduce a field multiplet whose components describe all electrons and then retain only the invariant term. This yields an action~\eqref{0211a} which coincides with the Hartree--Fock approximation in the semiclassical limit. Unfortunately the definition of the Green function given below~\eqref{0402a} is only formal. Hence it is not immediately clear how one could use this formalism to calculate the vacuum polarization.

From M.~Gell-Mann and F.~Low we know that the parameters of a quantum theory are functions of energy scale~\cite{gellmann} and the action obeys the Wilson--Polchinski equation~\cite{pol}. Above a sufficiently large infrared cut-off the Hartree--Fock action can be calculated using the theory of perturbation, i.e. the well-known  Uehling potential yields a quite accurate description at distances below one Compton length. At larger distances the Uehling potential is exponentially small. On the contrary all atomic orbitals are mainly above one Compton length. Consequently the long-range asymptote of the effective action is important.

As an approximation to the vacuum polarization in atoms one often uses the long-range potential of one-electron atom calculated by  Wichmann and Kroll~\cite{wk}. However it is still unclear what is the relevance of this approach to many-body systems. Indeed each Hartree--Fock system has a unique set of solutions which produces a distinct finite and a common divergent terms. One could also hope to calculate the Green function numerically by solving the Hartree--Fock equation and then summing up these solutions. However finding even a very small subset with an appropriate accuracy is a difficult computational task.

There are few observations behind the approximation which we propose here. For a Dirac--Coulomb system the $s$-orbitals form the leading contribution to the nonperturbative component of the vacuum polarization. Therefore for Hartree--Fock systems  we narrow our attention to $s$-orbitals. Although it is difficult to find a large number of solutions numerically it is possible to calculate a few eigenvalues corresponding to $s$-orbitals with lower principal quantum numbers~\cite{dirac19}, for Uranium see appendix~\ref{0611b}. For heavy elements we shall consider these values as a dilatation of the original Dirac--Coulomb spectrum. Then we replace the renormalization flow of the vacuum polarization by a piece-wise linear dilatation of the original flow corresponding to the Dirac--Coulomb polarization~\eqref{1307c}. In other words we calculate the vacuum polarization of a deformed Dirac--Coulomb operator provided we know its spectrum. Using this method the vacuum polarization in Uranium is one half of what one would obtain from the result of Wichmann and Kroll by simply scaling the charge of nuclei~\eqref{0610a}. In a narrow sense this statement is our main result. More broadly we describe here an approximation that one could apply for a variety of heavy elements if he knows how to find a sufficient number of eigenvalues. For a single-electron ion our numerical calculation is in an agreement with the exact solution~\cite{wk}. The computation and numerical analysis are performed using Mathematica~\cite{math12}.

Our approach is only valid for large electronic systems. It would be a mistake to apply the present method for lithium-like ions regardless the charge of nuclei. The reason for this is the $1/N$-expansion. Since $n$-point vertex functions are of order $N^{1-\frac{n}{2}}$ this expansion becomes meaningless for small~$N$, i.e. for atoms with few electrons. Hopefully for ions and atoms with only few electrons there exist other methods~\cite{indelicato}. Precisely our result should be understood as the leading term of $1/N$-expansion.

The accuracy of our numerical calculations is bounded above by the remainder of the Taylor expansion and an artifact which is a set of oscillations due to a finite UV cut-off~\eqref{2208a}. The only way to decrease these oscillations is to increase the UV cut-off. However, for a fixed radial distance a larger momentum cut-off would require larger values of the quantum number~$\kappa$. Unfortunately larger values of~$\kappa$ deteriorate the performance and lead to severe numerical errors in Mathematica. Thus a straightforward way to improve the precision of our result is to use a different numerical library.

\subsection{A short tour of the Dirac theory}
To introduce the relativistic propagator it is necessary to clarify the physical meaning of the negative energy solutions of the Dirac equation. In contemporary physics the particle--hole hypothesis has been replaced by the particle--antiparticle formalism. Hence it is appropriate to start with a short exposition of this theory. The reader familiar with QED should rather go to section~\ref{0807a}. The essence of the idea is a decomposition of the Klein--Gordon equation using quaternions
\begin{align}
i(\partial_0 + \sigma_i \partial_i)u&=  m v,&(\partial^2_0  - \nabla^2 + m^2) u&=0,\label{501a}\\
i(\partial_0 - \sigma_i \partial_i)v&=  m u,&(\partial^2_0  - \nabla^2 + m^2) v&=0,\label{501b}
\end{align}
where $\sigma_i$ are the Pauli matrices. If we define a metric $E=i \sigma_2$ in the linear space of two-component spinors then for any $2 \times 2$ complex matrix $\Lambda \in SL(2,\mathbb{C})$ with $\det \Lambda=1$ the bilinear form $\xi^\nu E_{\nu \mu} \eta^\mu$ is invariant under the transformation $\xi^\prime= \Lambda \xi$, $\eta^\prime = \Lambda \eta$. Clearly
\begin{align}
  \xi E \eta &= \det \|\xi \eta\|,&\|\xi \eta\|&=\begin{pmatrix} \xi_1& \eta_1\\ \xi_2 & \eta_2 \end{pmatrix},& \det \|\xi^\prime \eta^\prime\| &=\det (\Lambda \|\xi \eta\|)=\det \Lambda \det \|\xi \eta\|.
\end{align}
In the space of anti-hermitian matrices $M^{\lambda \dot{\mu}}$
\begin{equation}
  M=i\begin{pmatrix} x^0 + x^3 & x^1 - ix^2 \\ x^1 + ix^2 & x^0 -x^3 \end{pmatrix}= i(x^0 + x^i \sigma_i) \label{1204a}
\end{equation}
the transformation $M \mapsto M^{\prime}=\Lambda M \Lambda^+$ furnishes an irreducible representation of the proper orthochronous Lorentz group (the restricted Lorentz group):
\begin{align}
  \det M^{\prime}&=x^{\prime \mu} g_{\mu \nu} x^{\prime \nu},\\
  \det M&= x^\mu g_{\mu \nu} x^\nu,&\det M^{\prime}=\det M & \implies \exists L \in SO^{+}(1,3) : x^{\prime}=L x.
\end{align}
Here $x^{\mu}$ is a contravariant vector in Minkowski space with the metric $g_{\mu \nu}=(-1,1,1,1)$.
Van der Waerden proposed the dot symbol to mark the components whose transformation law uses the complex conjugate matrix, i.e. $\xi^{\prime \dot{\mu}}=\Lambda^* \xi^{\dot{\mu}}$. For each element of the restricted Lorentz group we have two matrices $\pm \Lambda$, in other words $SO^+(1,3) \sim SL(2,\mathbb{C})/Z_2$. There are two distinct elements of $SL(2,\mathbb{C})$ which are mapped to a single point in~$SO^+(1,3)$. A curve between these elements in $SL(2,\mathbb{C})$ is a loop in $SO^+(1,3)$. Clearly we can not continuously shrink this loop to the point, i.e. the group~$SO^+(1,3)$ is not simply connected. This is similar to non-relativistic quantum mechanics, a rotation by angle~$2 \pi n, n \in \mathbb{Z}$ which is a closed path from the identity to the identity transformation in $SO(3) \subset SO^+(1,3)$ corresponds to the sign factor $(\pm 1)^n$ of spinors, $SO(3) \sim SU(2)/Z_2$. Furthermore  it is easy to show that the  matrix $M_{\dot{\mu} \lambda}=(E M^* E^T)_{\dot{\mu} \lambda}$ with covariant indices corresponds to the covariant vector $x_\mu=g_{\mu \nu} x^\nu$. Since the partial derivatives transforms as a covariant vector we can write equations~\eqref{501a} and~\eqref{501b} in a form which is explicitly invariant under the spinor transformations
\begin{align}
i \partial_{\dot{\mu} \lambda}u^\lambda -  m v_{\dot{\mu}}&=0,&i\partial^{\lambda \dot{\mu}} v_{\dot{\mu}} -  m u^\lambda&=0,\label{501e}
\end{align}
where $\partial^{\lambda \dot{\mu}}=E^{\lambda \sigma} E^{\dot{\mu} \dot{\nu}} \partial_{\dot{\nu} \sigma}$. The definition of~$\partial_{\dot{\nu} \sigma}$ follows literally from equation~\eqref{1204a}. We complex conjugate the equation on the right and define a new spinor $w^\mu=v^{* \mu}$
\begin{align}
i \partial_{\dot{\mu} \lambda}u^\lambda  -  m w^*_{\dot{\mu}}&=0,&i\partial_{\dot{\mu} \lambda} w^\lambda -  m u^*_{\dot{\mu}}&=0. \label{501c}
\end{align}
With an auxiliary potential $A_{\dot{\mu} \lambda}$ these equations become $U(1)$ gauge invariant 
\begin{align}
i (\partial_{\dot{\mu} \lambda} -ie A_{\dot{\mu} \lambda})u^\lambda  -  m w^*_{\dot{\mu}}&=0,&i(\partial_{\dot{\mu} \lambda} + ie A_{\dot{\mu} \lambda}) w^\lambda -  m u^*_{\dot{\mu}}&=0.\label{901a}
\end{align}
The gauge transformation has the form
\begin{align}
u &\to e^{i e\alpha}u,&w &\to e^{-i e\alpha} w,&A_{\dot{\mu} \lambda} \to A_{\dot{\mu} \lambda} + \partial_{\dot{\mu} \lambda} \alpha.
\end{align}
It is clear from equations~\eqref{901a} that the spinors $u$ and $w$ describe particles of opposite charge. Let $b_{\dot{\mu}}=u^*_{\dot{\mu}}$ and complex conjugate the left equation in~\eqref{501c}
\begin{align}
i \partial^{\lambda \dot{\mu}}b_{\dot{\mu}}  -  m w^\lambda&=0,&i\partial_{\dot{\mu} \lambda} w^\lambda -  m b_{\dot{\mu}}&=0. \label{501d}
\end{align}
In the Weyl or chiral representation of the $\gamma$ matrices~\eqref{1204b} the equations~\eqref{501a} and~\eqref{501b} yield the Dirac equation
\begin{align}
(\slashed{\partial} + m) \psi&=0,&\psi&=\begin{pmatrix} u^\lambda\\v_{\dot{\mu}} \end{pmatrix},&\slashed{\partial}&=\gamma^\mu \partial_\mu\;.
\end{align}
Comparing \eqref{501d} and \eqref{501e} we see that the Dirac equation is invariant under the charge conjugation~$C$,
\begin{align}
C:  \begin{pmatrix}u\\v\end{pmatrix}&\to \begin{pmatrix}w\\b\end{pmatrix}=\begin{pmatrix}-i \sigma_2 v^*\\i \sigma_2 u^*\end{pmatrix},&\psi &\to \gamma^2 \psi^*,
\end{align}
the time reversal~$T$ and the parity~$P$ transformations,
\begin{align}
  T: \psi_{t,\vec{x}} &\to \gamma^1 \gamma^3 \psi^*_{-t,\vec{x}},& P: \psi_{t,\vec{x}} &\to i\gamma^0 \psi_{t,-\vec{x}}. \label{1602a}
\end{align}
Using invariance of equations~\eqref{501a}, \eqref{501b} under the space inversion~$P$ it is easy to see that a solution for~$v$ is the mirror image of~$u$. Since $v$ is the charge conjugate of $w$ it is natural to identify $u$ and $v$ with the chiral components of a particle. In the Dirac representation of the $\gamma$-matrices one obtains a different picture, more precisely, one of the components becomes dominant in the non-relativistic limit, i.e. for $E>0$ it corresponds to an electron. However in the Electroweak theory it is more convenient to keep $u$ and~$v$ independent because, as we all know after the Wu experiment, the chiral components interact differently. For the four component spinor~$\psi$ a positive energy solution describes the chiral components of a particle and, due to the $CPT$ symmetry, a negative energy solution which can not propagate forwards in time corresponds to the mirror image of the antiparticle. In relativistic quantum theory the time direction outside the light cone depends on a system of coordinates hence the both solutions are necessary. For $t<0$ the antiparticle propagates~$\psi_t=-e^{- i (- E) |t|} \psi_0$ with a positive energy~$-E$, see "The theory of positrons"~\cite{dic} by R. Feynman where he also proposed that the virtual pairs in the vacuum may interact with the real external charges.
\subsection{The Dirac density matrix}
We have seen in the previous section that the Dirac theory is a many-body theory. Consequently a probabilistic interpretation of the wave function becomes less obvious. Instead one should consider the transition probability amplitude from one state~$\psi_i$ to another~$\psi_j$, i.e. the scalar product between these states~$\langle \psi_i, \psi_j \rangle_{\mathcal{H}}$. In this sense the kernel of the transition matrix is the density matrix. For sake of simplicity we continue to consider only the case with no external field. Trying to give a probabilistic interpretation of the relativistic wave function Dirac proposed the following definition of the density matrix~\cite{dir},
\begin{equation}
  R_1(x,y) = P_-(x,y) -  P_+(x,y), \label{1911a}
\end{equation}
where $P_\pm$ are the projectors on the positive and negative spectrum respectively, i.e. $P_\pm\mathcal{H}=\mathcal{H}_\pm $. To find these projectors we consider Fourier transform of the Dirac equation
\begin{align}
(i \slashed{p} + m)\hat{\psi}(p)&=0,&\psi(x)&=\int \frac{d^4 p}{(2 \pi)^4}  e^{ipx} \hat{\psi}(p)\;.
\end{align}
Multiplying the both sides by $(-i \slashed{p} + m)$ one finds that $\hat{\psi}$ satisfies the Klein--Gordon equation
\begin{align}
(p^2 + m^2)\hat{\psi}(p)&=0,&\hat{\psi}(p)&= \hat{\phi}(\vec{p},\mathrm{sign}(p^0)) \delta(p^2 + m^2),
\end{align}
where $\hat{\phi}:\mathbb{R}^3 \times \mathbb{Z}_2 \to \mathbb{C}^4$ is a function on the mass shell. Since we restrict the functional space to solutions of the Klein--Gordon equation it is more convenient to use differently renormalized functions $\hat{\psi}_\pm(\vec{p})$ instead of $\hat{\phi}(\vec{p}, \pm)$
\begin{align}
\psi_{\pm}(x)&=\int \frac{d^3 p}{(2 \pi)^{\frac{3}{2}}} e^{\pm ipx} \hat{\psi}_{\pm}(\vec{p}),&\psi_\pm(x)&=(\mathcal{F}^{-1}_\pm \hat{\psi}_\pm)(x)\,,\\
\hat{\psi}_{\pm}(\vec{p})&=\int \frac{d^3 x}{(2 \pi)^{\frac{3}{2}}} e^{\mp ipx} \psi_{\pm}(x),&\hat{\psi}_\pm(\vec{p})&=(\mathcal{F}_\pm \psi_\pm)(\vec{p})\,.
\end{align}
Here $p^0=E$, $E=\sqrt{\vec{p}^{\, 2} + m^2}$, $\hat{\psi}_\pm \in \mathcal{H}_\pm$. Let $\psi \in \mathcal{H}_+ \oplus \mathcal{H}_-$ and apply the Fourier transformations
\begin{align}
(\mathcal{F}_+ \psi)(\vec{p})&=\hat{\psi}_+(\vec{p}) + e^{2i E t} \hat{\psi}_-(-\vec{p}),\label{1602e}\\
(\mathcal{F}_- \psi)(\vec{p})&=e^{-2i E t}  \hat{\psi}_+(-\vec{p}) + \hat{\psi}_-(\vec{p})\label{1602f}.
\end{align}
The components~$\psi_\pm$ satisfy the Dirac equation
\begin{align}
(i\slashed{p} + m)\hat{\psi}_{+}(\vec{p})&=0,&(i\slashed{p} - m)\hat{\psi}_{-}(\vec{p})&=0.\label{1602b}
\end{align}
Similar to~\eqref{1602a} the parity transformation gives
\begin{align}
(i\slashed{p} + m)\beta \,\hat{\psi}_{+}(-\vec{p})&=0,&(i\slashed{p} - m)\beta \,\hat{\psi}_{-}(-\vec{p})&=0,&\beta&=i\gamma^0. \label{1602c}
\end{align}
Using~\eqref{1602b} we obtain
\begin{align}
(i\slashed{p} - m)\beta \,\hat{\psi}_{+}(\vec{p})&= ((i\slashed{p} - m)\beta  + \beta (i\slashed{p} + m) )\hat{\psi}_{+}(\vec{p})= i\{\beta,\slashed{p}\}\hat{\psi}_{+}(\vec{p})=-2 E \hat{\psi}_{+}(\vec{p}),\\
(i\slashed{p} + m)\beta \,\hat{\psi}_{-}(\vec{p})&= i \{p,\beta\} \hat{\psi}_{-}(\vec{p})=-2E\hat{\psi}_{-}(\vec{p})\,. \label{1602d}
\end{align}
We define auxiliary projectors
\begin{equation}
\hat{P}_\pm(\vec{p})=\frac{-i\slashed{p} \pm m}{2E} \beta.
\end{equation}
Then from identities~\eqref{1602c}-\eqref{1602d} it follows the following properties
\begin{align}
\hat{P}_+(\vec{p}) \hat{\psi}_{+}(\vec{p})&=\hat{\psi}_{+}(\vec{p}),&\hat{P}_- (\vec{p}) \hat{\psi}_{-}(\vec{p})&=\hat{\psi}_{-}(\vec{p}),\\
\hat{P}_+(\vec{p}) \hat{\psi}_{-}(-\vec{p})&=0,&\hat{P}_-(\vec{p}) \hat{\psi}_{+}(-\vec{p})&=0.
\end{align}
Moreover $\hat{P}^2_\pm(\vec{p})=\hat{P}_\pm(\vec{p})$ and $\hat{P}_\pm(\vec{p}) \hat{P}_\mp(-\vec{p})  = 0$. Using~\eqref{1602e} and \eqref{1602f} we get $\hat{P}_\pm \mathcal{F}_\pm \psi=\hat{\psi}_\pm$ and thus $P_\pm=\mathcal{F}^{-1}_\pm \hat{P}_\pm \mathcal{F}_\pm$
\begin{align}
  P_+(x,y)&=\int \frac{d^3 p}{(2 \pi)^3} \frac{-i \slashed{p} + m}{2 p^0}\beta e^{ip(x-y)}=(-\slashed{\partial}_x+ m)\beta\int \frac{d^3 p}{(2 \pi)^3 2 p^0} e^{ip(x-y)},\\
  P_-(x,y)&=\int \frac{d^3 p}{(2 \pi)^3} \frac{-i \slashed{p} - m}{2p^0}\beta e^{-ip(x-y)} =-(-\slashed{\partial}_x+ m)\beta\int \frac{d^3 p}{(2 \pi)^3 2 p^0} e^{-ip(x-y)}.
\end{align}
Finally we have $P_\pm(x,y)=P_\pm(x-y)$,
\begin{align}
P_+(x)+P_-(x)&=(-\slashed{\partial}+ m)\beta \int \frac{d^3 p}{(2 \pi)^3 2 p^0} (e^{ipx}-e^{-ipx})=(-\slashed{\partial}+ m)\beta\Delta(x),\label{1802a}\\
P_+(x)-P_-(x)&=(-\slashed{\partial}+ m)\beta \int \frac{d^3 p}{(2 \pi)^3 2 p^0} (e^{ipx}+e^{-ipx})=(-\slashed{\partial}+ m)\beta\Delta_1(x).\label{1602g}
\end{align}
Here $\Delta_1$ is the Hadamard function, knows also as the even solution of the Klein--Gordon equation. The odd solution~$\Delta$ is the Pauli--Jordan function. This function vanishes for space-like intervals and is used in the quantum theory of fields to define the anticommutator $\{\psi_x, \psi^+_y\}=(-\slashed{\partial}+ m)\beta\Delta(x-y)$. Here $\psi_x$ is a generalized operator in the Hilbert space~$\mathcal{H}$. On the contrary $R_1=P_- - P_+$ produces non-local correlations. A detailed analysis of their singularities can be found in the original paper by Dirac~\cite{dir}. In non-relativistic quantum mechanics the expectation of an observable~$O$ is defined as
\begin{align}
\langle O \rangle_{\mathcal{H}} &= \int \limits_{(\vec{x},\vec{y}) \in \mathbb{R}^3 \times \mathbb{R}^3}  \psi^*(\vec{x}) O(\vec{x},\vec{y})\psi(\vec{y}),&(O \psi)(\vec{x})&= \int \limits_{\vec{y} \in \mathbb{R}^3}  O(\vec{x},\vec{y}) \psi(\vec{y}).
\end{align}
The simplest form of the density matrix is $\rho(\vec{y},\vec{x})=\psi(\vec{y})\psi^*(\vec{x})$ and the expectation is given by the trace of the observable with the density matrix, i.e.~$\langle O \rangle_{\mathcal{H}}=Tr (O \rho)$. The relativistic extension of the density matrix should give the propagator. Clearly $R_1$ allows propagation of particles with a negative energy forward in time. In the quantum theory of fields there are two equivalent definitions of the propagator~\cite{sw},
\begin{align}
  -i \hbar \Delta_F(x,y)&=\langle T \psi_x \psi^+_y \rangle_{\mathcal{H}}\,,\label{2602b}\\
  -i \hbar \Delta_F(x,y)&= \theta(x^0-y^0) P_+(x,y) - \theta(y^0-x^0)P_-(x,y) \,. \label{2602a}
\end{align}
For any space-like intervals
\begin{equation}
  \lim \limits_{t \to 0\pm} i \hbar \Delta_F(t,\vec{x})= \frac{1}{2}(R_1 \big|_{t=0} \mp \delta^3_{\vec{x}})=\frac{1}{2} R_1 \big|_{t=0}. \label{0507a}
\end{equation}

In 1927 J.~von Neumann defined the density~$\rho$ and the expectation of an observable~$\mathcal{O}$ for a canonical ensemble at a temperature~$T$, see~\cite{vanhove1958},
\begin{align}
\rho &= \frac{e^{-\beta H} }{Tr (e^{-\beta H})}\;,&\langle \mathcal{O} \rangle_\rho&=Tr( \mathcal{O} \rho),&\beta&=\frac{1}{\kappa T}.\label{0602a}
\end{align}
This definition has been subsequently extended to the grand canonical ensemble~\cite{ruelle,gljf}. Using this probability measure one obtains the 2-point Schwinger function, i.e. the covariance. There is a relation between the statistical field theory in Euclidean space and the quantum theory of fields in Minkowski space. In particular, one can obtain the Feynman propagator~$\Delta_F$ by performing the Wick rotation of the covariance~\cite{ow}. 
\section{The Hartree--Fock action}\label{0807a}
A usual way to formulate QED is the functional integral formalism, see e.g. L.D. Faddeev~\cite{fad}.
The fundamental formula for constructing Euclidean field theory, see e.g.~\cite{gljf}, is
\begin{align}
  \langle F, e^{-H \tau} G \rangle_{\mathcal{H}}&= \langle \theta F, T(\tau) G \rangle_{d \nu},& \langle  F, G \rangle_{d \nu} &= \int d \nu_{\Phi} F^*(\Phi) G(\Phi).
\end{align}
Here $F,G \in \mathcal{E}_+=\mathrm{span}\{e^{i\Phi(f)}: f \in \mathcal{D}_+\}$ are functionals on the set of test functions $\mathcal{D}_+=C^\infty_0(\mathbb{R}_+, \mathbb{R}^3)$ supported on the positive half of the time-axis; $\mathcal{H}=\overline{\mathcal{E}_+/\mathcal{N}}$ is the Hilbert space where $\mathcal{N}$ is the null space, i.e. $\mathcal{N}=\{A \in \mathcal{E}_+:\langle A, A \rangle=0\}$; $\theta$~and~$T(\tau)$ are the time reflection and time translation where an action~$\varepsilon$ of Euclidean group acts via  $(\varepsilon F)(\Phi)=F(\varepsilon \Phi)$, $(\varepsilon \Phi)(f)=\Phi(\varepsilon f)$. On the left hand side we have the inner product in the Hilbert space~$\mathcal{H}$, on the right -- the probability expectation. As usual we denote by $\int d \nu_\Phi$ a functional integral with a Gaussian measure on~$\mathcal{D}^\prime(\mathbb{R}^4)$
\begin{equation}
  e^{\frac{1}{2} \langle j C j\rangle }= \int d \nu_{A} \; e^{\langle j A \rangle} \label{2911a}. 
\end{equation}
The construction of Euclidean QED in the $\xi$-gauge leads to a formal path integral in~$\mathcal{D}^\prime(\mathbb{R}^4)$ with the Lagrangian semiclassical density which includes a gauge fixing term
\begin{equation}
  \Gamma_0=\frac{1}{4}  F^2_{\mu \nu} + \frac{1}{2 \xi}  (\partial_\mu A_\mu)^2.\label{1506a}
\end{equation}
The Feynman gauge corresponds to~$\xi=1$. Calculating the inverse of this bilinear form yields the covariance matrix
\begin{align}
C_{\mu \nu}(x)&=\int \frac{d^4p}{(2 \pi)^4}e^{ipx} \hat{C}_{\mu \nu}(p),&  \hat{C}_{\mu \nu}(p)&=\frac{\delta_{\mu \nu}}{p^2}\;.\label{2103a}
\end{align}
Let~$\mathcal{G}(\mathbb{R}^4)$ denote an infinite dimensional algebra whose generators are maps from~$\mathbb{R}^4$ to a Grassmann algebra~$\mathcal{G}$. A system with fermions is usually quantized using the Berezin integral~\cite{ber,sam}
\begin{equation}
e^{\langle \bar{\eta} S_0 \eta  \rangle}  = \int d \nu_{\bar{\psi}\psi} \; e^{\langle \bar{\eta} \psi \rangle  +  \langle \bar{\psi} \eta \rangle}.\label{2002a}
\end{equation}
Here $\bar{\eta},\eta \in \mathcal{G}(\mathbb{R}^4)$ are smooth function with compact support and $\bar{\psi},\psi \in \mathcal{G}^\prime(\mathbb{R}^4)$. One defines this integral for a finite dimensional Grassmann algebra and then makes sense of it in the limit. For Dirac fermions $\Gamma_0=\bar{\psi}(\slashed{\partial} + m)\psi$ and thus the covariance matrix is
\begin{align}
  S_0(x)&=\int \frac{d^4 p}{(2 \pi)^4} \hat{S}_0(p) e^{ipx},& \hat{S}_0(p)= \frac{-i \slashed{p} + m}{p^2 + m^2}\;.\label{2002b}
\end{align}
Here $\slashed{p}=p_\mu \gamma_\mu$ and $\{\gamma_\nu,\gamma_\mu\}=2\delta_{\mu \nu}$, $\gamma_0=\beta$.

To introduce interaction one substitutes $\slashed{\partial}$ with the covariant derivative wrt local gauge transformations, i.e. $\slashed{\partial} \to \slashed{\partial} - ie \slashed{A}$, and define a generating functional
\begin{align}
  Z_{qed}(K)=\int d \nu_\Phi \; e^{-L(\Phi) + \langle  K \one_\eta \Phi \rangle},&&L_0(\Phi)&= -ie\int  \bar{\psi}\slashed{A} \psi,\label{0407a}\\
\Phi \in (A_\nu,\bar{\psi}, \psi),\quad K \in (j_\nu, \eta, \bar{\eta}),&& (\one_k)_{\phi \bar{\phi}}&= \left\{\begin{matrix} -1 & k=\phi=\bar{\phi},\\\delta_{\phi \bar{\phi}}&\text{otherwise}.\end{matrix}\right.
\end{align}
In the perturbation theory $L$ is understood as a formal expansion over the Planck constant, i.e. the loop expansion $L=\sum_{l \geqslant 0} \hbar^lL_l$. The terms $L_{l}$ with $l>0$ are the counterterms.

In Quantum Mechanics the Hartree--Fock method is an approximation to the quantum many-body problem. One assumes that quantum mechanical $N$-particle state is an antisymmetric combination of one-particle wave functions~$\{\varphi_i \in \mathrm{H}^1\}$ known as the Slater determinant~\cite{slater}. The equations of motion are obtained by the variational principle, i.e. they correspond to the stationary points of the energy under the constraint that the $L_2$-norm of the wave functions be one. This leads to the Euler--Lagrange equation~$\h_i \varphi_i=\varepsilon_i \varphi_i$ where $\{\h_i\}^N_{i=1}$ are one-particle Hamiltonians. Usually one defines the Fock operator~$F$, substitutes one-particle Hamiltonians~$\h_i$ with the Fock operator~$F$ and solves the eigenvalue problem $F \varphi_i=\varepsilon_i \varphi_i$. For neutral and positively charged atoms and ions of~$N+1$ electrons the minimiser of the of the Hartree--Fock energy exists~\cite{simon}, and there are infinitely many stationary points of the same non-linear problem with~$N$ electrons~\cite{lewin}. The discrete spectrum of the Dirac Hamiltonian with a potential~$V$ such that for $r \to \infty$ it behaves as $V= C r^{\alpha -2}$ where $0<\alpha<2$ is also infinite~\cite{kurb}. The one-particle Hamiltonians~$\h_i$ are of this kind, see below~\eqref{2107a}.

To account for the vacuum polarization it is necessary to add to the method arbitrary number of virtual particles, i.e. we need to make a transition from the representation in the configuration space to a new canonical variable which describes the occupation numbers of quantum states~\cite{dir27}. A further development of this idea led V.~Fock to the introduction of the quantized wave function~$\psi$ and subsequently to a new form of the non-relativistic energy operator~\cite{fock}
\begin{equation}
  H=\int \limits_{x \in \mathbb{R}^3} \psi^+_x \left(\frac{- \Delta}{2 m} - \frac{e^2}{|x|}\right) \psi_x + \frac{e^2}{2} \int \limits_{x,y \in \mathbb{R}^3} \frac{\psi^+_x \psi^+_y \psi_y \psi_x}{|x-y|}\,.\label{0407b}
\end{equation}  
Here the first term is the Hamiltonian of an electron in a Coulomb potential and the second term gives two-particle interaction. This form can be obtained using the generating functional given above in~\eqref{0407a} if one integrates out all electromagnetic quantum fluctuations. Indeed given an external current~$\mathcal{J}$ and denoting the corresponding background field by~$\mathcal{A}$, i.e. $\mathcal{A}=C\mathcal{J}$, we obtain
\begin{align}
Z_{qed}(\mathcal{J},\eta,\bar{\eta})&=e^{\frac{1}{2}\langle \mathcal{J} \mathcal{A} \rangle}e^{W(\eta,\bar{\eta})},&e^{W(\eta,\bar{\eta})}&=\int d \nu_{\bar{\psi} \psi} e^{-L_{\mathcal{A}} + \langle \bar{\eta} \psi \rangle + \langle \bar{\psi} \eta \rangle   },
\end{align}
where
\begin{equation}
  L_{\mathcal{A}}= - ie \langle \bar{\psi} \slashed{\mathcal{A}} \psi \rangle + \frac{e^2}{2} \langle \bar{\psi} \gamma^\mu \psi C_{\mu \nu} \bar{\psi} \gamma^\nu \psi\rangle \label{1309f}.
  \end{equation}
For simplicity we restrict our discussion to the semiclassical approximation and deliberately omit all the counterterms. Let define the classical fields as usual
\begin{align}
\phi&=\frac{\delta W}{\delta \bar{\eta}},&\bar{\phi}&=-\frac{\delta W}{\delta \eta}.\label{1309b}
\end{align}
Each of these vacuum expectations has a non-vanishing value only in the presence of the corresponding external source. Using integration by parts~\eqref{0603a} and taking the semiclassical limit, i.e. $\hbar \to 0$, we obtain
\begin{align}
  (\slashed{\partial} + m)\phi &= -\frac{\delta L_{\mathcal{A}}(\bar{\phi},\phi)}{ \delta \bar{\phi}} + \eta ,&\bar{\phi} (\slashed{\partial} + m) &= \frac{\delta L_{\mathcal{A}}(\bar{\phi},\phi)}{ \delta \phi} + \bar{\eta} . \label{1309c} 
\end{align}
Solutions of these equations make stationary the following effective action
\begin{equation}
\Gamma_0=\langle \bar{\phi}(\slashed{\partial} -ie \slashed{\mathcal{A}} + m) \phi \rangle + \frac{e^2}{2} \langle  \bar{\phi} \gamma^\mu \phi C_{\mu \nu} \bar{\phi} \gamma^\nu \phi \rangle - \langle \bar{\phi} \eta \rangle - \langle \bar{\eta} \phi \rangle.
\end{equation}
Defining a quantum state in the Hilbert space as the Slater determinant and closely following V. Fock~\cite{fock} we would get a relativistic generalization of the Hartree--Fock equation~\eqref{0407b}.

Let perform a shift of the Gaussian measure~$d \nu_A$~\eqref{2911a} by a background field~$A_\mu \to A_\mu + \mathcal{A}_\mu$ 
\begin{align}
  Z_{qed}(K)&= e^{\frac{1}{2} \langle \mathcal{A} C^{-1} \mathcal{A} \rangle + \langle \mathcal{A}(j - \mathcal{J})\rangle + W(j- \mathcal{J},\bar{\eta}, \eta)},\\
  e^{W(K)}&=\int d \nu_\Phi \; e^{ - L(A + \mathcal{A}, \bar{\psi}, \psi) + \langle K \one_\eta \Phi \rangle}\;.
\end{align}
It is also convenient to define a measure $d \mu_{\bar{\psi} \psi}$
\begin{equation}
  e^{ \langle \bar{\eta} S \eta \rangle}=\int d \mu_{\bar{\psi} \psi} e^{ \langle \bar{\eta} \psi \rangle + \langle \bar{\psi} \eta \rangle }, \label{0503a}
\end{equation}
where $S$ is a solution of a non-homogeneous Dirac equation with an external potential~$\mathcal{A}$, e.g. a Coulomb potential,
\begin{equation}
  (\slashed{\partial} - i e \slashed{\mathcal{A}} + m) S_{xy}= \delta^4_{xy}. \label{0409b}
\end{equation}
The covariance~$S$ obtained by performing the Wick rotation of the propagator~$S^{\mathcal{M}}$ from Minkowski space. To uniquely define the propagator~$S^{\mathcal{M}}$ we shall also specify a walk around the poles. 

It is not immediately clear how one can write the partition function for an atom. However for an ion with a strong charge~$Z$ and a small number of electrons~$N+1$ the eigenvalues of the Dirac--Coulomb operator are close enough to the eigenvalues of the Fock operator and thus we can use the theory of perturbation at the first order in the two-particle interaction~\eqref{1309f}. A nonperturbative approach will be discussed below in section~\ref{0208a}. In Slater's approximation a $N$-particle state is the determinant
\begin{align}
\Phi_N&=\prod \limits^N_{i=1} \psi^+(f_i)  \,,& \langle \Phi_N, \Phi_N \rangle_{\mathcal{H}}&=1,&f_i&=\delta_t \varphi_i,& \langle \varphi^+_i \varphi_j \rangle&=\delta_{ij}.
\end{align}
Here $\varphi_i \in H^1(\mathbb{R}^3, \mathbb{C}^4)$ are wave functions of electrons. The state is invariant under the transformations of a symmetry group, i.e. $\varphi \mapsto g\varphi$ where~$g \in SU(N)$. We consider the 2-point correlation function of the last electron on the background given by the remaining electrons~$\Phi_N$. The functions~$\{\varphi_i\}_{i \in 1 \dots N}$ are eigenvectors of the Dirac--Coulomb Hamiltonian
\begin{align}
\bar{g}^f(x,y)&=\langle \Phi_N, \psi_{x} \,\psi^+_{y} \Phi_N \rangle_{\mathcal{H}},&(x,y) &\in (\mathbb{R}^4, \mathbb{R}^4).
\end{align}
\begin{figure} 
  \centering
\includegraphics[width=0.5\textwidth]{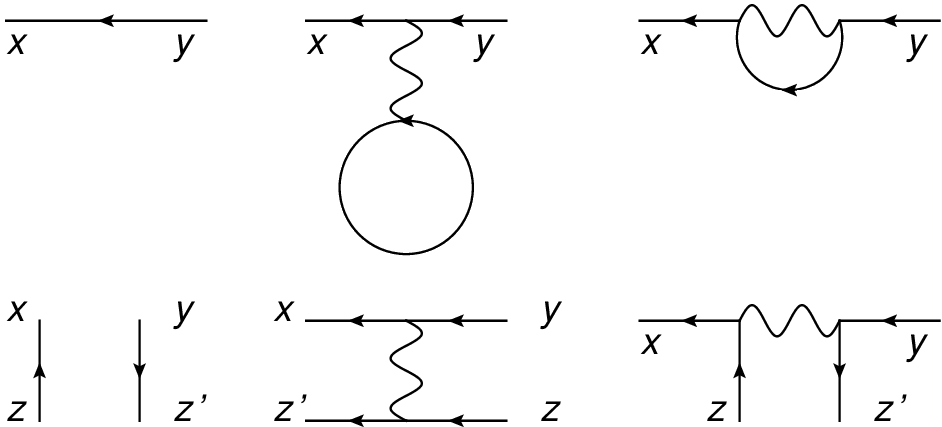}
\caption{$2(n+1)$-point correlator $\bar{g}$, $n \in \{0,1\}$.}  \label{0409a}
\end{figure}
It is not difficult to realize that $2(n+1)$-point functions~$\langle\Phi_0 \psi_{\vec{z}^{\, \prime}_1} \dots \psi^+_{\vec{z}_n}\Phi_0 \rangle$, $\Phi_0=1$ with two points connected to~$x$, $y$ and with the remaining $2n$ points connected to~$\varphi_i(\vec{z}_i)$, $\varphi^+_i(\vec{z}^{\,\prime}_i)$ describe the screening of the Coulomb potential and the exchange of the electron with all other $n$~electrons. In Figure~\ref{0409a} the diagrams in the second line can be obtained from the corresponding terms in the first line by a suitable modification of the walk around the poles of the propagator~\eqref{0409b}, i.e. after performing the Wick rotation a new covariance~$S^f$ will already contain all diagrams shown in the second line
\begin{align}
  S(x,y) &\mapsto S^f(x,y)=S(x,y) + \sum \limits^N_{i=1} S_i(x,y), & S_i(x,y)&=-e^{- \omega_i (x^0 - y^0)} \varphi_i(\vec{x}) \bar{\varphi}_i (\vec{y})\,.\label{1309a}
\end{align}
However at higher orders in the theory of perturbation we should account for functions with four and more external lines connected to the background state. Here we do not need to develop the expansion beyond the first order. Performing the Wick rotation to Minkowski space $S^{\mathcal{E}}(it)=-i S^{\mathcal{M}}(t)$, restricting~$\bar{g}^f$ to the hyperplane~$t=0$ and retaining only the leading order we get the result which is similar to equation~\eqref{0507a} 
\begin{align}
i \hbar \bar{g}^f_0(\vec{x},\vec{y}) &= \frac{1}{2}R^f_1(\vec{x},\vec{y}),&R^f_1(\vec{x},\vec{y})&=P^f_-(\vec{x},\vec{y}) - P^f_+(\vec{x},\vec{y}),\label{0607a} \\
\bar{g}^f(\vec{x},\vec{y})&=\sum \limits_{l \geqslant 0} \hbar^l \bar{g}^f_l(\vec{x},\vec{y}),&P^f_\mp(\vec{x},\vec{y})&= P_\mp(\vec{x},\vec{y}) \pm \sum \limits^N_{i=1}  \varphi_i(\vec{x}) \varphi^+_i(\vec{y})\,.\label{0607b}
\end{align}
We come to a representation in which the other electrons define the Fermi energy level, i.e. a path around the poles of the Feynman propagator. It is clear that the connection between the background state and the poles of the propagator is a general property which is independent of QED.
 
We now obtain one-particle Hamiltonian~$\h$ without resorting to the Feynman diagrams in Figure~\ref{0409a}. Assuming the renormalizability of QED we construct a generating functional in a such way that the counterterms preserve the Lorentz invariance and satisfy the Ward identities
\begin{align}
  Z(K)&= \int d \mu_\Phi \; e^{ -L + \langle K \one_\eta  \Phi \rangle },\quad Z(0)=1, \\
\mathcal{L} &=-ie Z_2\bar{\psi}\slashed{A} \psi - Z_2\delta m \bar{\psi} \psi +(Z_2 -1) \bar{\psi}S^{-1} \psi \nonumber \\
&\quad +(Z_3 -1) A_\mu C^{-1}_{\mu \nu} \mathcal{A}_\nu+ \frac{(Z_3 - 1) }{2}A_\mu C^{-1}_{\mu \nu} A_\nu \,.\label{2603a}
\end{align}
Convention used here is clear from the scaling properties of the measure given in appendix~\ref{2503a}. Here we assume that the Ward identities can be restored and intentionally omit all other terms which break the gauge invariance. This may require a further clarification. A mathematically meaningful definition of this integral is rather hopeless without an ultraviolet cutoff. But the introduction of a such cutoff will break the gauge invariance and some non-physical terms as the photon mass and $A^4$ coupling will appear in our calculations. The latter term is finite but depends on a way we perform the integration, the former is divergent~\eqref{0907b}. In the course of the renormalization procedure we restore the gauge invariance by normalizing the photon mass to zero. The decomposition given in~\eqref{2603a} preserves the gauge invariance because it reduces to the multiplicative renormalization. To see this let assume that the constant $Z_i$ are finite and independent of~$\hbar$, then we shall restore the Planck constant in the Feynman--Kac integral and consider the semiclassical limit~$\hbar \to 0$
\begin{equation}
  e^{\frac{1}{\hbar} W(K)}= \int d \mu_\Phi \; e^{ - \frac{1}{\hbar}L + \frac{1}{\hbar} \langle K \one_\eta \Phi \rangle }.
\end{equation}
Define the corresponding classical fields
\begin{align}
  a&=\hbar \frac{\delta W}{\delta j},&\phi&=\hbar \frac{\delta W}{\delta \bar{\eta}},&\bar{\phi}&=-\hbar\frac{\delta W}{\delta \eta}.\label{0505a}
\end{align}
Using integration by parts~\eqref{0603a},~\eqref{0803c} we obtain equations of motion in the limit~$\hbar \to 0$ 
\begin{align}
  C^{-1} a + \frac{\delta L}{\delta a} -j&=0,&S^{-1} \phi + \frac{\delta L}{\delta \bar{\phi}} - \eta&=0,&\bar{\phi} S^{-1}  - \frac{\delta L}{\delta \phi} - \bar{\eta}&=0.\label{2603b}
\end{align}
It follows that the semiclassical action has the form
\begin{equation}
  \Gamma_0=\frac{1}{2} \langle a C^{-1} a \rangle  +  \langle \bar{\phi}S^{-1} \phi \rangle  + L - \langle ja \rangle- \langle \bar{\phi} \eta \rangle - \langle \bar{\eta} \phi \rangle.
\end{equation}
Substituting~$L$ with the decomposition given in~\eqref{2603a} we see that this expression is a rescaling of variables in the initial action
\begin{align}
  \Gamma_0&=\frac{Z_3}{2} \langle (a + \mathcal{A})C^{-1}(a + \mathcal{A})\rangle + Z_2 \langle \bar{\phi}(\slashed{\partial} - ie (\slashed{a} + \slashed{\mathcal{A}})+ m - \delta m) \phi \rangle  \nonumber\\ & \quad - \langle (j+ C^{-1} \mathcal{A})a\rangle - \langle \bar{\phi} \eta \rangle - \langle \bar{\eta} \phi \rangle - \frac{Z_3}{2} \langle\mathcal{A} C^{-1} \mathcal{A} \rangle.
\end{align}
The last term does not appear in equations of motion~\eqref{2603b} and can be omitted. In the following we put $Z_i=1 + O(\hbar)$.

We also need to give a few important definitions. The 2-point correlation function is
\begin{align}
  \langle \psi_x \bar{\psi}_y \rangle_{d \mu}&= -Z^{\bar{\eta} \eta },&Z^{\bar{\eta} \eta }&=\frac{\delta^2}{\delta \bar{\eta}_x  \delta \eta_y } Z (0).
\end{align}
The generating functional of connected Schwinger functions is related to the functional~$Z$ by a simple combinatorial relation
\begin{equation}
  e^W=Z.
\end{equation}
Since $\bar{\psi}$, $\psi$ are odd elements of $\mathcal{G}^\prime(\mathbb{R}^4)$ their expectations vanish, $\langle \bar{\psi} \rangle_{d \mu}=\langle \psi \rangle_{d \mu}=0$. Consequently we have $W^{\bar{\eta} \eta }=Z^{\bar{\eta} \eta }$, and thus the leading order of~$W^{\bar{\eta} \eta }$ in the loop expansion coincides with the covariance~$-S$. Using integration by parts~$\eqref{0603a}$ we obtain an identity
\begin{equation}
  \int d \mu_\Phi \, e^{-L} \langle \bar{\eta} \psi \rangle \langle \bar{\psi} \eta \rangle = \langle \bar{\eta} S \eta \rangle + \int d \mu_\Phi \, e^{-L} \left( \langle \bar{\eta} S \frac{\delta^2 L}{\delta \bar{\psi} \delta \psi} S \eta \rangle - \langle \bar{\eta} S \frac{\delta L}{\delta \bar{\psi}} \rangle \langle \frac{\delta L}{\delta \psi} S \eta \rangle \right). \label{0803a}
\end{equation}
The propagators at the endpoints suggest the introduction of the connected amputated Schwinger functions. We define its generating functional~$\lfn$
\begin{align}
  W(K)&=\frac{1}{2}\langle j C j\rangle + \langle \bar{\eta} S \eta \rangle   -\lfn(Cj,\bar{\eta}S,S \eta),\\
  e^{-\lfn(a,\bar{\phi},\phi)}&=\int d \mu_\Phi \; e^{-L(A +a, \bar{\psi} + \bar{\phi},\psi + \phi)}. \label{0803b}
\end{align}
In particular for the 2-point correlator all these definitions yield
\begin{equation}
  \langle \psi_x \bar{\psi}_y \rangle_{d \mu}=-Z^{\bar{\eta}(x) \eta(y) }=-W^{ \bar{\eta}(x) \eta(y)}=S_{xy}+S_{xx^\prime} \lfn^{\bar{\phi}(x^\prime) \phi(y^\prime)} S_{y^\prime y}. \label{1503a}
\end{equation}
Hence identity~\eqref{0803a} can be written in the form
\begin{equation}
  \lfn^{\bar{\phi} \phi}= \int d \mu_\Phi \, e^{-L} \left( \frac{\delta^2 L}{\delta \bar{\psi} \delta \psi} - \frac{\delta L}{\delta \bar{\psi}} \frac{\delta L}{\delta \psi}\right), \label{0803d}
\end{equation}
where
\begin{align}
  \frac{\delta L}{\delta \psi}&=Z_2 ie \bar{\psi} \slashed{A} + Z_2 \delta m \bar{\psi} - (Z_2 -1) \bar{\psi} S^{-1},\\
 \frac{\delta L}{\delta \bar{\psi}}&=- Z_2 ie \slashed{A} \psi - Z_2 \delta m \psi + (Z_2 -1) S^{-1} \psi,\\
\frac{\delta^2 L}{\delta \bar{\psi}_x \delta \psi_y}&= Z_2 i e \slashed{A} \delta_{xy} + Z_2 \delta m \delta_{xy} - (Z_2 -1 ) S^{-1} \delta_{xy} \,.
\end{align}
One can obtain this equation directly from definition~\eqref{0803b} using the fact that odd fermionic correlators vanish. Using integration by parts~\eqref{0803c}  we have
\begin{align}
   \int d \mu_\Phi \, e^{-L} \langle f A \rangle &= \int d \mu_\Phi \, e^{-L} \langle f \Pi \rangle ,\\
   \int d \mu_\Phi \, e^{-L} \langle f A \rangle  \langle A g \rangle &=\int d \mu_\Phi \, e^{-L} \left(\frac{1}{Z_3}\langle f C g \rangle + \langle f \Pi \rangle \langle \Pi g \rangle \right),
\end{align}
where
\begin{equation}
\Pi_\alpha=ie\frac{Z_2 }{Z_3} C_{\alpha \beta} \bar{\psi} \gamma^\beta \psi  - \frac{Z_3 -1}{Z_3} \mathcal{A}_\alpha\, .
\end{equation}
Substituting these identities into~\eqref{0803d} we obtain
\begin{multline}
  \lfn^{\bar{\phi}(x) \phi(y)} = ie Z_2  \delta^4_{xy}  \langle \slashed{\Pi} \rangle_{d \mu} - \frac{Z^2_2 e^2}{Z_3} \langle \gamma^\nu \psi_x C_{\nu \mu}(x,y) \bar{\psi}_y \gamma^\mu \rangle_{d \mu}+ Z_2 \delta m \delta_{xy} - (Z_2 -1)S^{-1} \delta^4_{xy}\\
+\langle (Z_2 ie \slashed{\Pi} \psi + Z_2 \delta m \psi - (Z_2 -1) S^{-1} \psi)_x (Z_2 ie \bar{\psi} \slashed{\Pi}  + Z_2 \delta m \bar{\psi} - (Z_2 -1) \bar{\psi} S^{-1} )_y \rangle_{d \mu}\,,
\end{multline}
where the last term is a correction of the second order in the Planck constant. Retaining only the leading order we get
\begin{align}
  \lfn^{\bar{\phi}(x) \phi(y)}_1&= e^2 \delta^4_{xy} \gamma^\nu C_{\nu \mu}  Tr (S \gamma^\mu) - ie (Z_3 -1) \delta^4_{xy} \slashed{\mathcal{A}}  \nonumber\\
  &\quad - e^2 \gamma^\nu C_{\nu \mu}(x,y)S(x,y) \gamma^\mu +  \delta m \delta^4_{xy} - (Z_2 -1)S^{-1} \delta^4_{xy}\,.\label{1503b}
\end{align}
Each of constants $Z_2, Z_3, \delta m$ has divergent and finite parts. The finite part is fixed by the renormalization conditions. Retaining in equation~\eqref{1503a} the first order in the Planck constant we obtain the Dirac equation
\begin{equation}
 (\slashed{\partial} - ie \slashed{\mathcal{A}} + m - \lfn^{\bar{\phi} \phi}_1)W^{\bar{\eta} \eta }= - \delta^4\,. \label{0605a}
\end{equation}
Expression on the right hand in~\eqref{1503b} can be written in a more concise form using 1PI functions. The generating functional of 1PI functions is the effective action~$\Gamma$ which is defined by the Legendre transform~\eqref{0505a}
\begin{equation}
\Gamma(\Phi)=\langle K \one_\eta \Phi \rangle - W(K). \label{0410b}
\end{equation}
From this definition one can derive the following identity
\begin{equation}
  \frac{\delta^2 \Gamma}{\delta \Phi \delta \Phi} \one_{\bar{\eta}}  \frac{\delta^2 W}{\delta K \delta K} \one_\eta = \delta^4 \,. \label{0410c}
\end{equation}
We substitute into this equation the expansion of~$\Gamma$ over the Planck constant
\begin{equation}
  \Gamma(\Phi)= \frac{1}{2} \langle A C^{-1} A \rangle + \langle \bar{\psi} (\slashed{\partial} -ie (\slashed{\mathcal{A}} + \slashed{A}) + m) \psi \rangle + \sum \limits_{l \geqslant 1} \hbar^l \Gamma_l(\Phi),
\end{equation}
restrict it to $K=0$ and then retain only first order corrections. This yields
\begin{equation}
  (\slashed{\partial} - ie \slashed{\mathcal{A}} + m -\Sigma_{1} + ie \slashed{C} \Gamma^A_{1}  )  W^{ \bar{\eta}  \eta}= -\delta^4 \,.
\end{equation}
Subtracting equation~\eqref{0605a} we obtain
\begin{equation}
   \lfn^{\bar{\phi} \phi}_1=\Sigma_{1} - ie \slashed{C} \Gamma^A_{1} \,. 
\end{equation}

Let perform the Wick rotation back to Minkowski space-time. Recall that the Schwinger functions~$\mathcal{S}_n$ is a restriction of an analytic extension of the Wightman functions~$\mathcal{W}_n$ to Euclidean points~\cite{ow}
\begin{align}
  \mathcal{R}&: x \mapsto \begin{pmatrix}-i&0\\0&1\end{pmatrix}x,\label{1907a}\\
  \mathcal{R}_* &: \mathcal{S}_n(x_1, \dots x_n)=\mathcal{W}_n(\mathcal{R} x_1, \dots \mathcal{R}x_n), \quad x^0_j>x^0_{j+1}.
\end{align}
The covariance matrix of the vector field~$A$ transforms respectively under $SO(4)$ and $SO(1,3)$. For $x^0_1> x^0_2$ let $-i C^{\mathcal{M}}=\mathcal{W}_{AA}=\langle A A \rangle_{\mathcal{H}}$ and $C^{\mathcal{E}}=\mathcal{S}_{AA}$. Consequently we have
\begin{equation}
  \mathcal{R}_*:  iC^{\mathcal{E}}(x_1,x_2)=\mathcal{R} C^{\mathcal{M}}(\mathcal{R} x_1,\mathcal{R} x_2) \mathcal{R} \, .
\end{equation}
Indeed for $t>0$
\begin{equation}
C^{\mathcal{E}}(t,\vec{x})= \int \frac{d^3 p}{(2 \pi)^3} \frac{e^{-|\vec{p}| t + i\vec{p}\vec{x}} }{2 |\vec{p}|} \delta_{\mu \nu},
\end{equation}
and thus
\begin{align}
\mathcal{R}^{-1}C^{\mathcal{E}}(\mathcal{R}^{-1}x) \mathcal{R}^{-1}&= -i \int \frac{d^4 p}{(2 \pi)^4} e^{ipx} \hat{C}^{\mathcal{M}}_{\mu \nu}(p),&\hat{C}^{\mathcal{M}}_{\mu \nu}(p)&= \frac{g_{\mu \nu}}{p^2 - i \epsilon}.
\end{align}
The propagator in the Coulomb gauge can be obtained using a gauge transformation~\cite{dau}
\begin{align}
\hat{C}^{\mathcal{M}}_{\mu \nu}(p)&= \frac{g_{\mu \nu} + p_\mu f_\nu(p) + p_\nu f_\mu(p)}{p^2 - i \epsilon},&f_i&=-\frac{p_i}{2 \vec{p}^{\,2}},&f_0&=\frac{p_0}{2\vec{p}^{\,2}}.
\end{align}
This gives the following form
\begin{align}
\hat{C}^{\mathcal{M}}(p)&=\frac{1}{p^2 - i \epsilon} \begin{pmatrix}-\frac{p^2}{\vec{p}^{\,2}}& 0\\0& \delta_{ij}- \frac{p_i p_j}{\vec{p}^{\,2}}\end{pmatrix} + \xi \frac{p_\mu p_\nu}{(\vec{p}^{\, 2})^2}\,,& \epsilon&\to 0,&\xi &\to 0,
\end{align}
where $\xi$ is a device to obtain the inverse operator, see e.g.~\eqref{1506a}. Similarly for $x^0_1> x^0_2$ let $-iS^{\mathcal{M}}=\mathcal{W}_{\psi \bar{\psi}}=\langle \psi \bar{\psi}\rangle_{\mathcal{H}}$ and $S^{\mathcal{E}}=\mathcal{S}_{\psi \bar{\psi}}$. Applying the Wick rotation to the covariance of the spinor fields in~\eqref{2002b} we have
\begin{align}
  S^{\mathcal{E}}_0(\mathcal{R}^{-1} x) &= -i S^{\mathcal{M}}_0(x),& \hat{S}^{\mathcal{M}}_0(p)&=\frac{-i \slashed{p} + m}{p^2 + m^2 -i \epsilon}\;.
\end{align}
Let $\lfn^{\mathcal{E} \, \bar{\phi}(\mathcal{R}^{-1}x) \phi(\mathcal{R}^{-1}y)}=\lfn^{\mathcal{M} \, \bar{\phi}(x) \phi(y)}$. Then equation~\eqref{1503b} has the form
\begin{align}
  i\lfn^{\mathcal{M} \,\bar{\phi}(x) \phi(y)}_1 &= e^2 \delta^4_{xy}  \gamma^\mu C^{\mathcal{M}}_{\mu \nu}   Tr(-i S^{\mathcal{M}} \gamma^\nu) + ie (Z_3 -1 )\delta^4_{xy} \slashed{\mathcal{A}}  \nonumber\\
  &\quad -e^2  C^{\mathcal{M}}_{\mu \nu}(x,y) \gamma^\mu (-i S^{\mathcal{M}}(x,y)) \gamma^\nu  + \delta m \delta^4_{xy} - (Z_2 -1 )S^{-1} \delta^4_{xy} \label{1903a}
\end{align}
The divergent constants calculated by several authors, see e.g.~\cite{adk,sw}
\begin{align}
Z_2&=1 - \frac{\alpha}{4 \pi} D_2,&Z_3&=1 - \frac{\alpha}{3 \pi} D_3,&\delta m&=m \frac{3 \alpha }{4 \pi}  D_m,&D_a&=\log \frac{\Lambda^2_0}{c_a m^2},&\alpha&=\frac{e^2}{4 \pi},
\end{align}
where $\Lambda_0$ is an ultraviolet cut-off in the momentum space and $\{c_a\}$ are some finite constants. Using equation~\eqref{1503a} we obtain that the 2-point Green function~$-i g^f(x,y)=\mathcal{W}_{\psi  \bar{\psi}}(x,y)$ satisfies the Dirac equation with first order radiative corrections,
\begin{equation}
  (\slashed{\partial} - ie \slashed{\mathcal{A}} + m - i\lfn^{\bar{\phi} \phi}_1)g^f=\delta^4.\label{1107a}
\end{equation}
Since the Green function $S$ coincides with~$g^f$ at the leading order in the loop expansion we can substitute~$S$ with~$g^f$ in equation~\eqref{1903a}. 
Furthermore because the quantities are invariant under the time translation we can perform the Fourier transform
\begin{align}
g^f_\omega &= \int dt \, e^{i \omega t} g^f(t),& \lfn^{\bar{\phi} \phi}_\omega &= \int dt \, e^{i \omega t} \lfn^{\bar{\phi} \phi}_1(t).
\end{align}
Put $V_\omega=-i\beta  \lfn^{\phi \bar{\phi}}_\omega$,  $V_\omega \in \mathcal{D}^\prime(\mathbb{R}^3,\mathbb{R}^3)$. From equation~\eqref{1107a} we read off that $g^f_\omega$ is the resolvent of the Hamiltonian
\begin{align}
  (\h- \omega)\bar{g}^f_\omega &= \delta^3,&\bar{g}^f&=g^f \beta,&\h&=\alpha^k (\partial_k-ie \mathcal{A}_k) -  e \mathcal{A}_0 + \beta m + V_\omega,\label{2105b}
\end{align}
where $\alpha^k=\beta \gamma^k$ are the Dirac $\alpha$-matrices. Here $g^f$ depends on the external potential~$\mathcal{A}$ which we shall restrict to the Coulomb potential of the nuclear charge $-Ze$. The resolvent can also be written using the bilinear formula
\begin{align}
  \bar{g}^f_\omega(\vec{x},\vec{y})&=\int d z \, \frac{ \varphi_z(\vec{x}) \varphi^+_z(\vec{y})}{z_\mp  -\omega},&z_\mp&=\left\{\begin{matrix}z - i\epsilon& z \in \sigma_+(\h),\\ z + i \epsilon & otherwise,\end{matrix}\right. & \h \varphi_z&= z \varphi_z,\label{2105a}
\end{align}
where $\sigma_+(\h)$ is the essential and discrete spectra of unoccupied orbitals. The position of the residues has been chosen such that $\lim \limits_{t \to 0 \pm} -i \bar{g}^f(t) =\pm P^f_\pm$, see also~\eqref{0607b}. 

Let us define the instantaneous density matrix~$\rho^f$ as the restriction of the Green function to~$t=0$, i.e. $\rho^f =  i  \bar{g}^f |_{t=0}$. Performing the inverse Fourier transform and applying the Cauchy contour integration theorem we have
\begin{equation}
\rho^f(\vec{x},\vec{y})=i\bar{g}^f(\vec{x},\vec{y})\Big|_{t=0} = i\int \frac{d \omega}{2 \pi} \bar{g}^f_\omega(\vec{x},\vec{y})=\frac{1}{2}R^f_1(\vec{x},\vec{y}),\label{2707b}
\end{equation}
\begin{equation}
R^f_1(\vec{x},\vec{y})=P^f_-(\vec{x},\vec{y}) - P^f_+(\vec{x},\vec{y}),\quad P^f_\pm(\vec{x},\vec{y})=\sum \limits_{\omega \in \sigma_\pm(\h)} \varphi_\omega(\vec{x}) \varphi^+_\omega(\vec{y}).
\end{equation}
Similar to equation~\eqref{0607a}  we have
\begin{equation}
\rho^f(\vec{x},\vec{y})=\rho(\vec{x},\vec{y}) + \sum \limits^N_{i=1} \varphi_i(\vec{x})\varphi^+_i(\vec{y})\,.
\end{equation}
Furthermore, because we have defined~$\bar{g}^f$ by putting the electron on a non-trivial Fermi energy level we also have
\begin{equation}
i\bar{g}^f(t,\vec{x},\vec{y})=i\bar{g}(t,\vec{x},\vec{y}) + \sum \limits^N_{i=1} e^{-i \omega_i t} \varphi_i(\vec{x})\varphi^+_i(\vec{y})\,.
\end{equation}
In the following we make a crude approximation by assuming that the external magnetic field vanishes and by omitting the magnetic interaction of electrons, i.e. we restrict the photon propagator to the Coulomb field
\begin{align}
  C_{\mu \nu}(t,r)&=\begin{pmatrix}-\frac{\delta_t}{4 \pi r}&0\\0&0\end{pmatrix},&\hat{C}_{\mu \nu}(p)&=\begin{pmatrix}-\frac{1}{p^2_l}&0\\0&0\end{pmatrix}. \label{1707a}
\end{align}
However to calculate the self-energy of electron one should use the complete photon propagator. Finally omitting the delta function~$\delta^3_{\vec{x} \vec{y}}$ in the diagonal terms the kernel of the Hamiltonian~\eqref{2105b} has the form
\begin{align}
\h_{\vec{x}\vec{y}}&=\alpha^k \partial_k  -\frac{\alpha Z}{r} + \beta m +   \alpha  \int \limits_{\vec{z} \in \mathbb{R}^3} \frac{Tr \sum \limits^N_{i=1}\varphi_i(\vec{z}) \varphi^+_i(\vec{z})}{|\vec{x} - \vec{z}|} - \alpha \frac{\sum \limits^N_{i=1}  \varphi_i(\vec{x}) \varphi^+_i(\vec{y})}{|\vec{x}-\vec{y}|}  \nonumber \\
&\quad -\frac{\alpha Z}{r}(Z_3 -1 )+ \alpha  \int \limits_{\vec{z} \in \mathbb{R}^3} \frac{Tr \rho(\vec{z},\vec{z})}{|\vec{x} - \vec{z}|} \nonumber\\
&\quad  +  (Z_2 -1)(\alpha^k \partial_k  +  m \beta -\frac{\alpha Z}{r} - \omega) - \delta m \beta   -  e^2 \beta (C_{\mu \nu} \gamma^\mu ig \gamma^\nu)_\omega  \, . \label{1911c}
\end{align}
The fourth and fifth terms are the direct and exchange terms respectively.

The above argument motivates us to define the Hartree--Fock approximation using the usual partition function but with a modified covariance. The additive terms to the covariance~\eqref{1309a} are equivalent to $2N$-dimensional Grassmann integral over a set of auxiliary variables~$\{\bar{\psi}_i,\psi_i\}^N_{i=1}$ with the covariance~$S_i$. The variables~$\bar{\psi}_i, \psi_i$ of each color have non-vanishing vacuum expectations~$\bar{\varphi}_i,\varphi_i$. The color can not be observed and there is only one particle of each color. 
\begin{definition}
We define a $SU(N)$ invariant generating functional of connected Schwinger functions as follows
\begin{align}
  e^{W(\veta,\bar{\veta})}&= \int  d \mu_{\bar{\vpsi} \vpsi} \, e^{-\frac{1}{2}  \langle (\bar{\psi}_{\dot{s}}  \gamma^\mu \psi_s) C_{\mu \nu} (\bar{\psi}_{\dot{r}} \gamma^\nu \psi_r) F^{\dot{s} \dot{r}}_{sr}\rangle  + \langle \bar{\psi}_i \eta_i \rangle + \langle \bar{\eta}_i \psi_i \rangle }\label{0211a}\\
F^{\dot{s}\dot{r}}_{sr}&= \frac{e^2}{N}( \delta_{\dot{s}s}\delta_{\dot{r}r} + \delta_{\dot{s}r}\delta_{\dot{r}s})\,.
\end{align}
\end{definition}
Here we omit the counterterms. The definition of the classical fields literally repeats expression~\eqref{1309b}. As above these fields satisfy the Euler--Lagrange equations~\eqref{1309c} with the following action
\begin{equation}
\Gamma_0=\langle \bar{\vphi}(\slashed{\partial} -ie \slashed{\mathcal{A}} + m) \vphi\rangle + \frac{e^2}{2 N} \sum \limits^N_{ij=1} \langle (\bar{\phi}_i \gamma^\mu \phi_i ) C_{\mu \nu} (\bar{\phi}_j \gamma^\nu \phi_j )  + ( \bar{\phi}_i \gamma^\mu \phi_j) C_{\mu \nu}  (\bar{\phi}_j \gamma^\nu \phi_i)  \rangle  \,.
\end{equation}
Omitting the magnetic interaction~\eqref{1707a} we obtain an identity $\hh \phi_a=\omega_a \phi_a$, 
\begin{align}
\hh_{\vec{x}\vec{y}}&=\alpha^k\partial_k - \frac{\alpha Z}{r} + \beta m + \alpha_N \int \limits_{\vec{z} \in \mathbb{R}^3} \frac{ Tr(-\phi_i \bar{\phi}_i \beta)_{\vec{z}} }{|\vec{x} - \vec{z}|} + \alpha_N \frac{ \phi_i(\vec{x}) \bar{\phi}_i(\vec{y}) \beta}{|\vec{x} - \vec{y}|},&\alpha_N&=\frac{\alpha}{N}.
\end{align}
The kernel of one-particle Hamiltonian is $\h_{\vec{x}\vec{y}}=\langle \Phi_{n-1}, \hh_{\vec{x}\vec{y}} \Phi_{n-1} \rangle$, i.e.
\begin{equation}
\h_{\vec{x}\vec{y}}=\alpha^k\partial_k - \frac{\alpha Z}{r} + \beta m + \alpha_N \int \limits_{\vec{z} \in \mathbb{R}^3} \frac{ \sum \limits_{i \neq a} Tr(\varphi^+_i \varphi_i )_{\vec{z}} }{|\vec{x} - \vec{z}|} - \alpha_N \frac{ \sum \limits_{i \neq a} \varphi_i(\vec{x}) \varphi^+_i(\vec{y})}{|\vec{x} - \vec{y}|}\,. \label{2107a}
\end{equation}

Usually one solves Dirac--Coulomb eigenvalue problem using the spherical coordinates system. In the Dirac representation of the $\gamma$-matrices the radial Dirac equation is
\begin{equation}
  - G^\prime + \frac{\kappa}{x}G= (z-1 +  \frac{\gamma}{x})F, \quad F^\prime + \frac{\kappa}{x}F=(z+1 +  \frac{\gamma}{x}) G,\label{2405a}
\end{equation}
\begin{align}
  z&=\frac{\lambda}{m},&\gamma&=Z \alpha,&x&=rm,&\kappa &\in \mathbb{Z} \backslash \{0\},&\varphi_\lambda(r)&=\frac{1}{r}\begin{pmatrix}F_r Y^m_{|\kappa+ \frac{1}{2}| - \frac{1}{2}} \\iG_r Y^m_{|\kappa - \frac{1}{2}| - \frac{1}{2}} \end{pmatrix} . \label{1307a}
\end{align}
Here $F,G$ are the radial wave functions and $Y^m_{j \pm \frac{1}{2}}$ are the spin spherical harmonics, for notation see e.g.~\cite{swain1,mess,shiff}. Note that these equations are invariant under the charge conjugation $F \leftrightarrow G$, $\gamma \leftrightarrow -\gamma$ $z \leftrightarrow -z$, $\kappa \leftrightarrow -\kappa$. For~$0<z_n<1$ there exists a discrete set of eigenvectors~$\{G_n, F_n\}$ which is dense in~$\mathrm{H}^1(\mathbb{R})$. For $-1<z\leqslant 0$ there are no solutions in~$\mathrm{H}^1(\mathbb{R})$, see~\cite{shiff}. For $|z|>1$ every solution is bounded for $x  \to \infty$~\cite{hl}.

\section{The Wilson--Polchinski equation} \label{0208a}
To account for radiative corrections one needs to add the vacuum polarization and self-energy terms to the Hamiltonian. In this section we write the renormalization group (RG) equation using an effective propagator~$G$ on the background corresponding to the atomic orbitals of the ground state. The calculation of radiative corrections on non-trivial backgrounds considered already by many authors, see e.g.~\cite{coleman,jackiw}. As an introduction to the RG theory the reader might consider the original review written by K. Wilson~\cite{wil}. There is an extensive list of references on the subject with applications to condensed matter physics~\cite{weg}, the field theory~\cite{pol,zin,wet,chk,riv} and stochastic pdes~\cite{kup}.

We restrict the solution to the leading contribution in the $1/N$-expansion~\cite{gn,wil2} which is a meaningful approximation for large atoms and molecules.  In the large-$N$ limit all $n$-point irreducible functions with $n>2$ are negligible in the vacuum~$N^{1 - \frac{n}{2}}$~\cite{hooft}. Only the 2-point vertex function is nonperturbative and should be computed using RG method. At the leading order in the $1/N$ expansion all one-particle Hamiltonians~$\h_i$ are equivalent. Let denote by~$\Lambda_0$ and $\Lambda$ the ultraviolet and infrared cut-offs which restrict the space of distributions on $\mathrm{H}^1$ to a subset
\begin{equation}
  \mathcal{D}^\prime_{\Lambda \Lambda_0}(\mathbb{R}^3)=\{\varphi_z \in \mathrm{H}^{-1} : \h \varphi_z = z \varphi_z ,\; |z^2-m^2|>\Lambda^2 \;, |z| \leqslant \Lambda_0 \}. \label{1807a}
\end{equation}
The generating functional~\eqref{0211a} with the both cut-offs has the form
\begin{equation}
  e^{W^{\Lambda \Lambda_0}(\veta,\bar{\veta})}= \int  d \mu^{\Lambda \Lambda_0}_{\bar{\vpsi} \vpsi} \, e^{-L^{\Lambda_0 \Lambda_0} + \langle \bar{\vpsi} \veta\rangle + \langle \bar{\veta} \vpsi \rangle },
\end{equation}  
where $L^{\Lambda_0 \Lambda_0}$ includes the counterterms. Since the Lagrangian is time translation invariant the Legendre transform~\eqref{1309b} gives the Hartree--Fock~$\mathcal{E}^{0 \Lambda_0}_{HF}$ energy multiplied by a time interval~$T$~\cite{coleman85}. Here we are interested in the limit $\Lambda \to 0$,
\begin{equation}
\Gamma^{0 \Lambda_0}(\bar{\vpsi}, \vpsi)=  T\mathcal{E}^{0 \Lambda_0}_{HF} - \langle \bar{\vpsi} \beta \boldsymbol{\omega} \vpsi\rangle= \langle \bar{\vpsi} \veta \rangle +  \langle \bar{\veta} \vpsi \rangle - W^{0 \Lambda_0}(\veta,\bar{\veta}),\label{1511b}
\end{equation}
where the orbital energies~$\boldsymbol{\omega}$ are Lagrange multipliers. Furthermore since our goal is an effective low energy theory we restrict generating functionals to the low energy components, e.g. $\veta,\bar{\veta} \in \mathcal{G}_{0 \Lambda}$. Using the background field method we translate the integration variables~$\vpsi \to \vpsi+\vphi$ by a field~$\vphi \in \mathcal{G}_{0 \Lambda}$ and then obtain
\begin{equation}
e^{W^{0 \Lambda_0}(\veta,\bar{\veta})}=e^{-\langle \bar{\vphi} S^{-1}_{0 \Lambda_0} \vphi\rangle + \langle \bar{\vphi} \veta \rangle +  \langle \bar{\veta} \vphi \rangle}\int  d \mu^{0 \Lambda_0}_{\bar{\vpsi} \vpsi}e^{-L^{\Lambda_0 \Lambda_0}(\bar{\vpsi} + \bar{\vphi},\vpsi + \vphi) + \langle \bar{\vpsi} (\veta - S^{-1}_{0 \Lambda_0} \vphi) \rangle +  \langle( \bar{\veta} - \bar{\vphi} S^{-1}_{0 \Lambda_0}) \vpsi \rangle}.
\end{equation}
Next we decompose the measure~$d \mu^{0 \Lambda_0}_{\bar{\vpsi} \vpsi}$ into $d \mu^{0 \Lambda}_{\bar{\vpsi}_1 \vpsi_1}d \mu^{\Lambda \Lambda_0}_{\bar{\vpsi}_2 \vpsi_2}$. Since $\vphi,\veta \in \mathcal{G}_{0 \Lambda}$ the integrand depends on~$\vpsi_2$ and~$\bar{\vpsi}_2$  only via the term $L^{\Lambda_0 \Lambda_0}(\bar{\vpsi}_1 + \bar{\vpsi}_2 + \bar{\vphi},\vpsi_1 + \vpsi_2+ \vphi)$. Consequently the integration over $d \mu^{\Lambda \Lambda_0}_{\bar{\vpsi}_2 \vpsi_2}$ yields the generating functional of connected amputated Schwinger functions with the infrared cut-off~$\lfn^{\Lambda \Lambda_0}$,
\begin{equation}
e^{W^{0 \Lambda_0}(\veta,\bar{\veta})}=e^{-\langle \bar{\vphi} S^{-1}_{0 \Lambda} \vphi\rangle + \langle \bar{\vphi} \veta \rangle +  \langle \bar{\veta} \vphi \rangle}\int  d \mu^{0 \Lambda}_{\bar{\vpsi} \vpsi}e^{-\lfn^{\Lambda \Lambda_0}(\bar{\vpsi} + \bar{\vphi},\vpsi + \vphi) + \langle \bar{\vpsi} (\veta - S^{-1}_{0 \Lambda} \vphi) \rangle +  \langle( \bar{\veta} - \bar{\vphi} S^{-1}_{0 \Lambda}) \vpsi \rangle}.
\end{equation}
The functional $\lfn^{\Lambda \Lambda_0}$ contains the radiative corrections of perturbative QED. By expanding $\lfn^{\Lambda \infty}$ at the point $\vphi$ we get
\begin{align}
e^{W^{0 \infty}(\veta,\bar{\veta})}&=e^{-I^{\Lambda}(\bar{\vphi},\vphi) + \langle \bar{\vphi} \veta \rangle +  \langle \bar{\veta} \vphi \rangle + W^{0 \Lambda}_\vphi (\veta - \delta_{\bar{\vphi}} I^{\Lambda}(\bar{\vphi}, \vphi),\bar{\veta} + \delta_{\vphi} I^{\Lambda}(\bar{\vphi},\vphi))},\label{1511a}\\
e^{W^{0 \Lambda}_\vphi(\veta,\bar{\veta})}&=\int  d \mu^{0 \Lambda}_{\bar{\vpsi} \vpsi} e^{- \sum \limits_{n \geqslant 2} \lfn^{\Lambda \infty}_n(\bar{\vpsi},\vpsi) + \langle \bar{\veta} \vpsi\rangle + \langle \bar{\vpsi},\veta \rangle},\label{2911b}\\
\lfn^{\Lambda \infty}_n(\bar{\vpsi},\vpsi)&=\sum \limits_{j+i=n} \frac{\left(\vpsi \delta_{\vphi}\right)^i \left(\bar{\vpsi} \delta_{\bar{\vphi}}\right)^j}{i!j!} \lfn^{\Lambda \infty}(\bar{\vphi},\vphi),\\
I^{\Lambda}(\bar{\vphi},\vphi)&=\langle \bar{\vphi} S^{-1}_{0 \Lambda} \vphi \rangle + \lfn^{\Lambda \infty}(\bar{\vphi},\vphi)\,.
\end{align}
The action~$I^{\Lambda}$ includes all radiative corrections above the infrared cut-off~$\Lambda$. The bilinear term~$\lfn^{\Lambda \infty}_2(\bar{\vpsi},\vpsi)$ can be decomposed in the form
\begin{equation}
   \lfn^{\Lambda \infty}_2(\bar{\vpsi},\vpsi)= \langle \bar{\vpsi} (V-\Sigma) \vpsi \rangle - \frac{1}{2}\langle \bar{\psi}_{\dot{s}} \bar{\psi}_{\dot{r}}  Q^{\dot{s}\dot{r}}   +   \bar{Q}_{sr} \psi_s \psi_r \rangle . \label{2702b}
\end{equation}
In the semiclassical approximation the auxiliary tensors are 
\begin{align}
V^{\dot{s}}_{s} &= F^{\dot{s}\dot{r}}_{sr}  \delta_{xy} \gamma^\mu    C_{\mu  \nu} Tr(-\phi_r \bar{\phi}_{\dot{r}} \gamma^\nu ),&Q^{\dot{r}\dot{s}}&= F^{\dot{s}\dot{r}}_{sr}  \,\gamma^\mu \phi_s(x) C_{\mu \nu}(x,y) \gamma^\nu \phi_r(y)\,,\label{2901a}\\
\Sigma^{\dot{s}}_{s} &=F^{\dot{s}\dot{r}}_{sr}  C_{\mu \nu}(x,y) \gamma^\mu  (-\phi_r(x)\bar{\phi}_{\dot{r}}(y)) \gamma^\nu,&\bar{Q}_{sr}&= F^{\dot{s}\dot{r}}_{sr}  \, \bar{\phi}_{\dot{s}}(x) \gamma^{\mu} C_{\mu \nu}(x,y)  \bar{\phi}_{\dot{r}}(y) \gamma^\nu\,.\label{2901b}
\end{align}
We can account for the bilinear form by defining a new measure~$d \tilde{\mu}_\Phi$
\begin{align}
  e^{\frac{1}{2}\langle K \one_{\eta} \g \one_{\bar{\eta}} K \rangle}&=\int d \tilde{\mu}_{\Phi}\, e^{\langle  K \one_\eta \Phi \rangle},&K&=(\veta,\bar{\veta}),&\Phi&=(\bar{\vpsi},\vpsi).\label{0302d}
\end{align}
Here~$\g$ is an effective propagator of the particles
\begin{align}
  \g&=\begin{pmatrix} -G^t \bar{Q} \tilde{G}& -\tilde{G}^t\\\tilde{G} & -\tilde{G} Q G^t\end{pmatrix},& \tilde{G}&=(1 + G Q G^t \bar{Q})^{-1}G, \label{0402a}
\end{align}
where $G$ is the Green function of the Hartree--Fock operator with the standard walk around the poles
\begin{equation}
  (\slashed{\partial} - ie \slashed{\mathcal{A}} +m + V-\Sigma)G= \delta^4_{xy}.\label{2702a}
\end{equation}
Using the measure~$d \tilde{\mu}_\Phi$ we can write
\begin{equation}
  \int d \mu^{0 \Lambda}_{\Phi}\, e^{-\lfn^{\Lambda \infty}_2(\bar{\vpsi},\vpsi)+ \langle  K \one_\eta \Phi \rangle}= \left(\frac{\det \s_{0 \Lambda}}{\det \g_{0 \Lambda}}\right)^{\frac{1}{2}}   \int d \tilde{\mu}^{0 \Lambda}_{\Phi} \, e^{\langle  K \one_\eta \Phi \rangle}.
\end{equation}
It follows that the generating functional~$W^{0 \Lambda}_\vphi$ has the form
\begin{align}
  e^{W^{0 \Lambda}_\vphi(\veta,\bar{\veta})}&=e^{\frac{1}{2} Tr \log  \g^{-1}_{0 \Lambda}\s_{0 \Lambda}} e^{\tilde{W}^{0 \Lambda}_\vphi(\veta,\bar{\veta})},\label{1711a}\\e^{\tilde{W}^{0 \Lambda}_\vphi(\veta,\bar{\veta})}&=\int d \tilde{\mu}^{0 \Lambda}_{\bar{\vpsi} \vpsi} \; e^{- \sum \limits_{n > 2} \lfn^{\Lambda \infty}_n(\bar{\vpsi},\vpsi) + \langle \bar{\veta} \vpsi\rangle + \langle \bar{\vpsi},\veta \rangle}.
\end{align}
Substituting~\eqref{1511a} into~\eqref{1511b} and denoting $\delta \bar{\vpsi} = \bar{\vpsi} - \bar{\vphi}$, $\delta \vpsi = \vpsi - \vphi$  we get an expansion
\begin{equation}
\Gamma^{0 \infty}(\bar{\vpsi}, \vpsi)= I^{\Lambda}(\bar{\vphi},\vphi) + \langle \delta \bar{\vpsi} \frac{\delta I^\Lambda(\bar{\vphi},\vphi)}{\delta \bar{\vphi}} \rangle + \langle \delta \vpsi  \frac{\delta I^\Lambda(\bar{\vphi},\vphi)}{\delta \vphi}  \rangle + \Gamma^{0 \Lambda}_\vphi(\delta \bar{\vpsi}, \delta \vpsi).\label{0302a}
\end{equation}
Here $\Gamma^{0 \Lambda}_\vphi$ is the Legendre transform of $W^{0 \Lambda}_\vphi$. Furthermore using equation~\eqref{1711a} we have
\begin{equation}
\Gamma^{0 \Lambda}_\vphi(\bar{\vpsi}, \vpsi)=-\frac{\hbar}{2} Tr \log  \g^{-1}_{0 \Lambda}\s_{0 \Lambda}+ \tilde{\Gamma}^{0 \Lambda}_\vphi(\bar{\vpsi}, \vpsi).\label{0302b}
\end{equation}
We choose the background fields to coincide with the classical fields, i.e. $\vphi=\vpsi$, $\bar{\vphi}=\bar{\vpsi}$ and
\begin{equation}
\Gamma^{0 \infty}(\bar{\vphi}, \vphi)=I^{\Lambda}(\bar{\vphi},\vphi)+\Gamma^{0 \Lambda}_\vphi(0).\label{1511c}
\end{equation}
Using~\eqref{0302b} we have
\begin{equation}
\Gamma^{0 \infty}(\bar{\vphi}, \vphi)=I^{\Lambda}(\bar{\vphi},\vphi) - \frac{\hbar}{2} Tr \log  (\g^{-1}_{0 \Lambda}  \s_{0 \Lambda}) + \tilde{\Gamma}^{0 \Lambda}_\vphi(0).\label{0312a}
\end{equation}
Since $\tilde{\Gamma}^{0 \Lambda}_\vphi(0)=O(\hbar^2)$ this equation gives the well known result of 1-loop approximation, see e.g.~\cite{jackiw,coleman}. The Hartree--Fock energy~$\mathcal{E}^{0\infty}_{HF}$ has the following form
\begin{equation}
\mathcal{E}^{0\infty}_{HF}(\bar{\vphi},\vphi)=\mathcal{E}_{HF}^{\Lambda \infty}(\bar{\vphi},\vphi) - \frac{\hbar}{2} Tr \log  (\g^{-1}_{0 \Lambda}  \s_{0 \Lambda})_{0\times \mathbb{R}^3} + O(\hbar^2).
\end{equation}
We have used the time translation symmetry, i.e. $I^\Lambda=T\mathcal{E}_{HF}^{\Lambda \infty}- \langle \bar{\vphi} \beta \boldsymbol{\omega} \vphi\rangle$. The Taylor expansion of the logarithm yields the perturbative result~\eqref{1911c}. In addition to the effective action~$\tilde{\Gamma}^{0 \Lambda}_\vphi$ we also define the corresponding reduced action~$\tilde{\Gamme}^{0 \Lambda}_\vphi$,
\begin{equation}
  \tilde{\Gamma}^{0 \Lambda}_\vphi(\Phi)= \frac{1}{2}\langle \Phi  \g^{-1}_{0 \Lambda}  \Phi \rangle + \tilde{\Gamme}^{0 \Lambda}_\vphi(\Phi).\label{1507a}
\end{equation}

Let $\lambda \in (0,\Lambda]$ be a floating infrared cut-off. Here~$\g_{\lambda \Lambda}$ is the propagator with the floating infrared cut-off, i.e. $\g_{\lambda \Lambda}=\sigma_{\lambda \Lambda} \g$ where $\sigma_{\lambda \Lambda}$ is the spectral projection of~$\h$ given in~\eqref{1807a}.  In the following we will omit the subscript and the tilde in the notation for the effective action~$\tilde{\Gamme}_\vphi$. Furthermore, $\dot{\g}_{\lambda \Lambda}$ is a shorthand for $\partial_\lambda \g_{\lambda \Lambda}$; $\Gamme^{\dprime}$ stands for $\Delta \Gamme$ where $\Delta_{ij} =\delta_{\Phi_i} \delta_{\Phi_j}$ and $\Phi_i \in (\bar{\vpsi},\vpsi)$; $\lfn^{\dprime}=\Delta \lfn$ where $\Delta_{ij}=\delta_{\Phi_i} \delta_{\Phi_j}$ and $\Phi_i \in (\bar{\vphi},\vphi)$. Let assume that ${\Gamma^{\dprime \lambda \Lambda}}^{-1}$ is a Neumann series expansion on the interval of energy scales~$(0,\Lambda]$
\begin{equation}
{\Gamma^{\dprime \lambda \Lambda}}^{-1}= -\sum \limits^\infty_{n=0}\left( \g_{\lambda \Lambda} \Gamme^{\dprime \lambda \Lambda}\right)^n \g_{\lambda \Lambda}\,.
\end{equation}
\begin{thm}The RG equation for the effective action has the form~\cite{mor,wet93}
\begin{equation}
\dot{\Gamme}^{\lambda \Lambda}=\frac{\hbar}{2}\langle  \dot{\g}_{\lambda \Lambda} \Gamme^{\dprime \lambda \Lambda} (1-\g_{\lambda \Lambda} \Gamme^{\dprime \lambda \Lambda})^{-1} \rangle.\label{0312b}
\end{equation}
\end{thm}
\emph{Proof.}
Using the definition of the Grassmann measure~\eqref{0302d} it is straightforward to obtain the derivative $d \dot{\mu}$ wrt the infrared cut-off~$\lambda$
\begin{equation}
  \int d \dot{\mu}_{\Phi} e^{ \langle K \one_{\eta} \Phi \rangle}= -\frac{\hbar}{2}\int d \mu_{\Phi} \langle \delta_{\Phi}  \dot{\g}   \delta_{\Phi} \rangle e^{ \langle K \one_{\eta} \Phi \rangle}.
\end{equation}
This yields the Wilson--Polchinski equation for the functional~$\lfn$ defined in~\eqref{0803b}
\begin{equation}
  \dot{\lfn}= \frac{\hbar}{2} \left(\langle  \dot{\g}   \lfn^\dprime \rangle + \langle \lfn^\prime   \dot{\g}   \lfn^\prime \rangle \right) . \label{0410a}
\end{equation}
The Legendre transform~\eqref{0410b} of the free energy $W(K)=\frac{1}{2} \langle K \one_{\eta} \g  \one_{\bar{\eta}} K \rangle - \lfn( \g  \one_{\bar{\eta}} K)$ gives
\begin{equation}
  \Gamme(\Psi)= \lfn(\Phi)-\frac{1}{2}\langle (\Psi - \Phi) \g^{-1} (\Psi -\Phi) \rangle.
\end{equation}
Applying the derivative~$\partial_\Lambda$ and using equation~\eqref{0410a} we have
\begin{equation}
\dot{\Gamme}= \frac{\hbar}{2} \langle  \dot{\g}  \lfn^\dprime \rangle. \label{0410d}
\end{equation}
Equation~\eqref{0410c} gives the identity
\begin{equation}
\lfn^\dprime =\Gamme^\dprime (1 -  \g   \Gamme^\dprime)^{-1}.
\end{equation}
Together with~\eqref{0410d} this yields the RG equation~\eqref{0312b}. \hfill$\blacksquare$\null

The reduced action $\Gamme$ is proportional to the fine-structure constant which is small. Furthermore the integration path~$(0,\Lambda]$ is short. Therefore we omit all nonlinear terms in the renormalization group equation~\eqref{0312b}. Finally we perform the Wick rotation to Minkowski space~\eqref{1907a}, i.e. $\Gamme^{\mathcal{E}}=i \Gamme^{\mathcal{M}}$.  At the leading order this gives us a flow equation
\begin{align}
\dot{\Gamme}^{\lambda \Lambda}&= \frac{\hbar}{2} \langle i\dot{\g}_{\lambda \Lambda} \Delta \rangle \Gamme^{\lambda \Lambda}\,,& \g_{\lambda \Lambda}&=\begin{pmatrix}0& -G^t_{\lambda \Lambda}\\G_{\lambda \Lambda}&0 \end{pmatrix}.
\end{align}
Using notations of~\eqref{2901a} and \eqref{2901b} in 1-loop approximation the flow equation has the form
\begin{equation}
\dot{\Gamme}^{\lambda \Lambda}= \hbar \langle i \dot{G}_{\lambda \Lambda}(V - \Sigma)\rangle.
\end{equation}
Clearly the equation is non-trivial only for 2-point function~$\Gamme^{\bar{\vpsi} \vpsi}= \delta_{\bar{\vpsi}}  \delta_{\vpsi}\Gamme(0)$. In the limit $N \to \infty$  only the diagonal terms of the color indices remain
\begin{equation}
\dot{\Gamme}^{\bar{\vpsi} \vpsi \, \lambda \Lambda}_{xy}=- \frac{\hbar  e^2 }{N} \left(\delta_{xy}\gamma^\nu C_{\mu \nu} Tr( i\dot{G}_{\lambda \Lambda} \gamma^\mu) -  \gamma^\mu i\dot{G}_{\lambda \Lambda}(x,y) C_{\mu \nu}(x,y) \gamma^\nu  \right). \label{1907b}
\end{equation}
\section{The vacuum polarization}
We will consider only the vacuum polarization, i.e. the first term in~\eqref{1907b}, and will also omit the magnetic interaction, see~\eqref{1707a}. It follows from definition~\eqref{1507a} that $\Gamme^{\bar{\vpsi} \vpsi}$  vanishes at zero order in~$\hbar$. At the leading order we make the following truncation of the effective action
\begin{equation} 
\Gamme^{\lambda \Lambda} =     T \langle \bar{\vpsi}_{\vec{x}} \,\alpha \beta \int_{\vec{z} \in \mathbb{R}^3} \frac{\hbar \rho^{\lambda \Lambda}_{\vec{z}}}{ |\vec{x}-\vec{z}|} \vpsi_{\vec{x}} \rangle_{0 \times \mathbb{R}^3}.\label{2307a}
\end{equation} 
Here $T$ is a finite time interval. It is convenient also to define the Dirac density matrix
\begin{equation}
\frac{1}{2}R^{\lambda \Lambda}_N =i G^{\lambda \Lambda}_{0\times \mathbb{R}^3,0\times \mathbb{R}^3} \beta.
\end{equation}
Substituting these definitions into equation~\eqref{1907b} and omitting the second term we obtain
\begin{equation}
\rho^{\lambda \Lambda}= \rho^{\lambda_1 \Lambda} - \sum \limits_{\{\lambda_n\}>\lambda}  \frac{1}{2N } \int \limits^{\lambda_n}_{\max(\lambda_{n+1},\lambda)}d \omega \,   Tr \dot{R}^{ \omega \Lambda}_N\,,\quad \forall \lambda \leqslant \lambda_1 .\label{2207e}
\end{equation}
Here the trace runs over the color and spinor indices. The vacuum polarization~$R_N$ becomes negligible at long distances. At shorter distances, up to five Compton lengths, the screening and exchange are small compared to the Coulomb potential. However there is a significant dilatation of the spectrum. For each spectral interval~$(\lambda_{n+1},\lambda_n]$, i.e. for each $x \in(\frac{1}{n+1},\frac{1}{n}]$, we replace the flow~$\partial_x R^{\omega \Lambda}_N=\dot{R}^{\omega \Lambda}_N \omega^\prime$ by a linear dilatation of the flow~$\partial_{x} R^{\gamma x \, \Lambda}_1$ corresponding to the Dirac--Coulomb operator,
\begin{align}
\partial_x R^{\omega \Lambda}_N &\mapsto \bigoplus \limits^N  \left( \frac{\lambda_n - \lambda_{n+1}}{\delta \omega_n}\partial_x R^{\gamma x(\omega)\, \Lambda}_1 \right),&\delta \omega_n&=\frac{\gamma}{n(n+1)},\label{1307c}\\
x(\omega)&=\frac{1}{n}\left(1 - \frac{1}{n+1}\frac{ \lambda_n - \omega}{\lambda_n - \lambda_{n+1}}\right),&\omega &\in (\lambda_{n+1},\lambda_n]\,.\label{2407b}
\end{align}
At very short distances, e.g. below one Compton length, the Uehling's induced charge~$U$ becomes dominant. It follows from~\eqref{1511a} that $\rho^{\Lambda \Lambda}=U^{\Lambda}$ assuming that~$\Lambda$ is sufficiently large, e.g. $\Lambda>5$. In the flow equation~\eqref{2207e} we omit the interval~$(\lambda_1,\Lambda]$. For $\lambda>\lambda_1$ the direct and exchange terms are small perturbation of the Coulomb potential. Therefore there should be only subleading corrections to the Wichmann and Kroll density. Primary we are interested in scales where the perturbative theory is not feasible. The nonperturbative component of the vacuum polarization is the difference between~$\rho$ and the Fourier transform of the Uehling's running coupling~\eqref{2007a}, 
\begin{align}
\nu(r)&=\rho(r)  - U(r)\,, \quad r\in \mathbb{R}^3_+\,, \quad |r|> 1\,,  \\
U(r)&=e^2 Z \int \frac{d^3 \vec{p}}{(2 \pi)^3} e^{-i pr} \pi(0,\vec{p})\,,& \int d^3 x\,  U(x)&=0\,.
\end{align}
The running coupling~$\pi(0,\cdot)$ is not a function in~$L_1(\mathbb{R}^3)$ and the Fourier transform does not exist in the usual sense. One needs to rescale the charge of nuclei~\eqref{1911c} by adding a counterterm and to use the renormalization condition above in order to extract a finite part. We should say that this renormalization condition only needed to define the nonperturbative component~$\nu$ which will add a finite shift to the charge of nuclei. Without this shift our renormalization procedure would be incorrect.

First we consider one electron in the potential of the Uranium nuclei, i.e. $Z=92$. To calculate~$R^{\Lambda \Lambda_0}_1$ we use definition of the density matrix~\eqref{1911a}, the spherical decomposition~\eqref{1307a} and the solutions of the radial Dirac equation given in appendix~\ref{1907d},
\begin{align}
\frac{1}{2}Tr\, R^{\Lambda \Lambda_0}_1(r)&= \frac{\yy^{\Lambda \Lambda_0}(|r|)}{8 \pi r^2}\,,&\yy^{\Lambda \Lambda_0}&=\sum \limits^{|\kappa|\leqslant K}_{\kappa \in \mathbb{Z}\backslash \{0\}}2 |\kappa| \big(\yy^{\Lambda \Lambda_0}_{\kappa -} - \yy^{\Lambda \Lambda_0}_{\kappa +} - \tilde{\yy}^{\Lambda}_{\kappa}  \big),\label{2207c}\\
\yy^{\Lambda \Lambda_0}_{\kappa \pm}&=\int \limits^{\Lambda_0}_\Lambda dp \, (|F^{\kappa p}_\pm|^2 + |G^{\kappa p}_\pm|^2),&\tilde{\yy}^{\Lambda}_{\kappa}&=\sum \limits^{|\kappa|+ j \leqslant M_\Lambda}_{\kappa j \in \substack{\mathbb{Z}_+ \mathbb{N}_+ \\ \mathbb{Z}_- \mathbb{N}} }( |\tilde{F}^{\kappa j}|^2 + |\tilde{G}^{\kappa j}|^2)\,.
\end{align}
The sum over~$\kappa$ is a finite sum of $2K$ terms where the radial cut-off~$K$ is fixed $K=55$. The last sum runs over the principal quantum numbers~$m=|\kappa| + j$  such that $p_m \geqslant \Lambda$ where $p_m$ is the corresponding momentum, for details see~\eqref{2107b}. Because of the spherical symmetry we use the radial scaling. For a fixed~$M_\Lambda$ we calculate the vacuum polarization~$\yy^{\Lambda \Lambda_0}$ with different cut-offs~$\Lambda_0$. Since the cut-offs $\Lambda_0$ and $K$ are finite we restrict the calculations to an appropriate bounded interval~$[a^{\Lambda_0},b^{\Lambda_0 K}]$. 

\begin{figure}[t]
\hfill \includegraphics[scale=0.55]{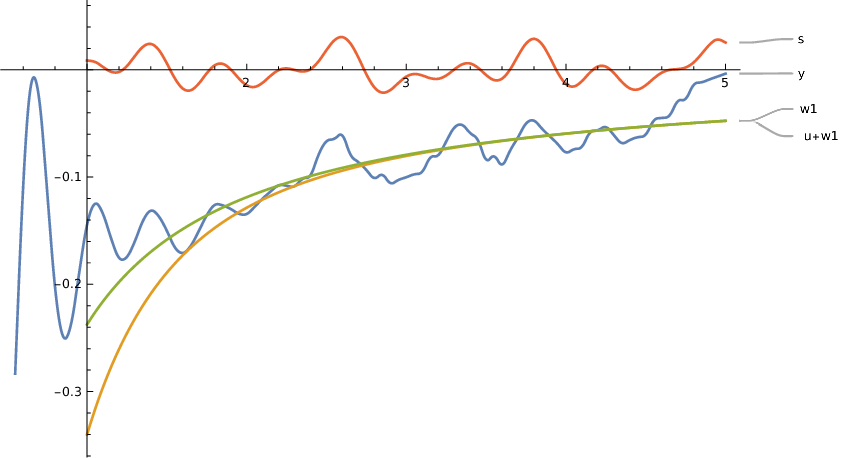}\hfill \includegraphics[scale=0.55]{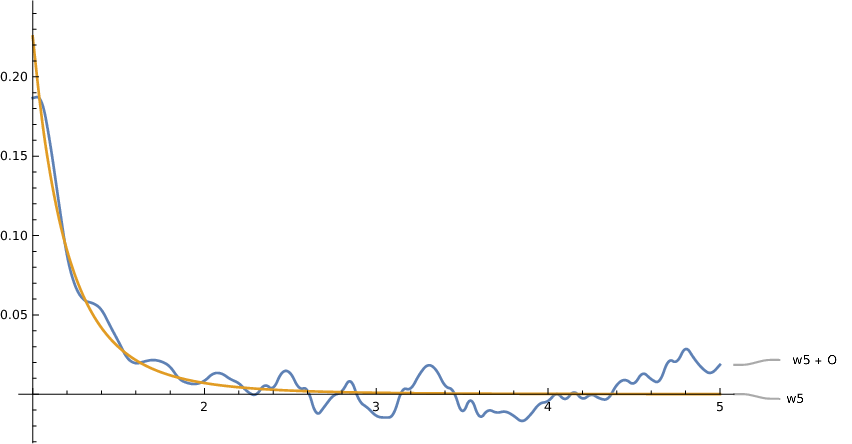}\hfill\null  
\caption{The decomposition of $\yy^{\Lambda \Lambda_0}$ with $Z=92$, $M_\Lambda=11$, $\Lambda_0=7.58$.}\label{2107c}
\end{figure}
Next we compute the Laplace transform~\eqref{2207b} and perform the following decomposition
\begin{equation}
\yy^{\Lambda \Lambda_0}(x)=\mathsf{c}^{\Lambda_0}_1 x +  \uu(x) +\mathsf{c}^{\Lambda_0}_2 \delta_x  + \frac{\ww^{\Lambda \Lambda_0}_1}{x}  + \sss^{\Lambda \Lambda_0}(x)+ \frac{\ww^{\Lambda \Lambda_0}_5}{x^5} + O_x,\label{2208a}
\end{equation}
\begin{align}
\uu(x)&=8 \pi x^2 U(x),&\sss^{\Lambda \Lambda_0}(x)&=\sum \limits_{\omega \in \Omega^{\Lambda \Lambda_0}} \mathrm{c}^{\Lambda \Lambda_0}_{\omega} e^{i \omega x}. \label{2207a}
\end{align}
The first component is divergent and must be subtracted, see also~\eqref{0907b}. The second and third components are the Uehling's charge density and a finite charge renormalization, respectively.  The fourth constant~$\ww^{\Lambda \Lambda_0}_1$ is a non-physical $A^4$ coupling, i.e. an artifact. This artifact depends on the way one defines the momentum integration and it vanishes in the limit $\Lambda \to 0$ and $\Lambda_0 \to \infty$. The sixth constant~$\ww^{\Lambda \Lambda_0}_5$ gives the nonperturbative vacuum polarization. The oscillations~$\sss^{\Lambda \Lambda_0}$ are another artifact appearing as a result of the finite cut-offs, i.e. $\Lambda$, $\Lambda_0$, $K$. The oscillations become larger outside the interval~$[a^{\Lambda_0},b^{\Lambda_0 K}]$. To calculate the frequencies~$\omega$ we perform the decomposition, restrict the Laplace transform to the imaginary axis, choose a neighborhood~$\mathcal{B}$ of the highest spike, then find the maximum~$\omega \in \mathcal{B}$ of the convolution~$\hat{O} * (c_1 \hat{ \sin} + c_2 \hat{\cos})_{\mathcal{B}}$ where $c^2_1 + c^2_2=1$. We add the frequencies~$\pm \omega$ to the set~$\Omega^{\Lambda \Lambda_0}$ and repeat the decomposition until the remainder~$O$ is sufficiently small. We perform these calculations with $10\%$ of accuracy. To measure the remainder and oscillations we define two norms
\begin{align} 
\|O\|^2_2&= \frac{1}{(b-a)^2 \hat{\ee}^2_4(0)}\int \limits^b_a dx \, O^2_x ,&\|\hat{\sss} \|^2_{\hat{2}}&= \frac{1}{\ee^2_4(0)}\sum \limits_{\omega \in \Omega_+} |\mathrm{c}^{\Lambda \Lambda_0}_{-\omega} \hat{\sss}_{- \omega}(0) + \mathrm{c}^{\Lambda \Lambda_0}_\omega \hat{\sss}_\omega(0)|^2.
\end{align} 
For a fixed upper bound of the remainder we choose a decomposition that minimizes the oscillations. An example of the decomposition is shown in Figure~\ref{2107c}. We omit the linear term in the figure. In the left panel we have the original~$\yy^{\Lambda \Lambda_0}$ which is calculated numerically~\eqref{2207c}, the oscillations~$\sss^{\Lambda \Lambda_0}$, the Uehling's density~$\uu$ and fourth component combined. In the right panel we have the last two terms of the decomposition. The oscillations vanish only in the limit $\Lambda \to 0$ and $\Lambda_0 \to \infty$. They do not vanish at a finite~$\Lambda>0$ even in the limit $\Lambda_0 \to \infty$.

In order to calculate the limits~$\{\ww^{\Lambda \infty}_5\}$ we define an objective function~$f$ and use the gradient descent to find the minimiser
\begin{equation}
f(\{\ww^{\Lambda \infty}_5\},\beta,\eta,)= \sum \limits_{\{\Lambda \Lambda_0 \} } (\ww^{\Lambda \infty}_5 + \beta  \Lambda^{- \eta}_0 - \ww^{\Lambda \Lambda_0}_5)^2.
\end{equation}
This yields $\beta=22.72$, $\eta=2.35$ and the limits~$\{\ww^{\Lambda \infty}_5\}$. To simplify the notation let us put $\ww_5(x)=\ww^{\gamma x\, \infty}_5$. Using a similar minimization method we find a piecewise differentiable approximation
\begin{align}
\ww_5(x)&=\upsilon \, x^{\xi_1}+ \chi\,,&x& \in (0, t_{1}],\\
\frac{\partial \log \ww_5(x)}{\partial \log x}&=\xi_i\,,&x& \in (t_{i-1}, t_{i}],
\end{align}
where
\begin{equation}\upsilon=0.72\,,\quad\chi=0.03\,,\quad \mbox{
\begin{tabular}{rrr}
$\xi_i$&$t_{i-1}$,&$t_{i}$\\
\hline
1.36&0&0.13\\
1.35&0.13&0.20\\
1.24&0.2&0.23\\
0.84&0.23&1
\end{tabular}}
\end{equation}
For $x>1$ the running constant $\ww_5(x)$ steeply approaches zero, i.e. $\ww_5^{\Lambda \infty}=0$. The running constant~$\nu_5^{\Lambda \infty}$ is generated at the scale of $1s$ orbital and then decreases as we integrate the flow~\eqref{2207e}, 
\begin{align}
\nu_5(x)&=  \ww_5(1) - \sum \limits^{x<x_n}_{n=1} \frac{\lambda_n - \lambda_{n+1}}{\delta \omega_n}\int \limits^{x_n}_{\max(x_{n+1},x)}dx  \, \ww^{\prime}_5\,,&\delta \omega_n&=\frac{\gamma}{n(n+1)},&x_n=&\frac{1}{n}.\label{2507c}
\end{align}
where $\nu_5(x(\omega))=\nu^{\omega \, \Lambda}_5$, see~\eqref{2407b}. Here we have substituted $\nu_5(1)$ with $\ww_5(1)$. The nonperturbative component at large distances is
\begin{equation}
\nu^{\lambda \Lambda}(r)=\frac{\nu^{\lambda \Lambda}_5}{8 \pi |r|^7}\,,\quad r \in \mathbb{R}^3_+\,, \quad |r|>1. \label{2507f}
\end{equation}

\begin{figure}[t]
\hfill \includegraphics[scale=0.55]{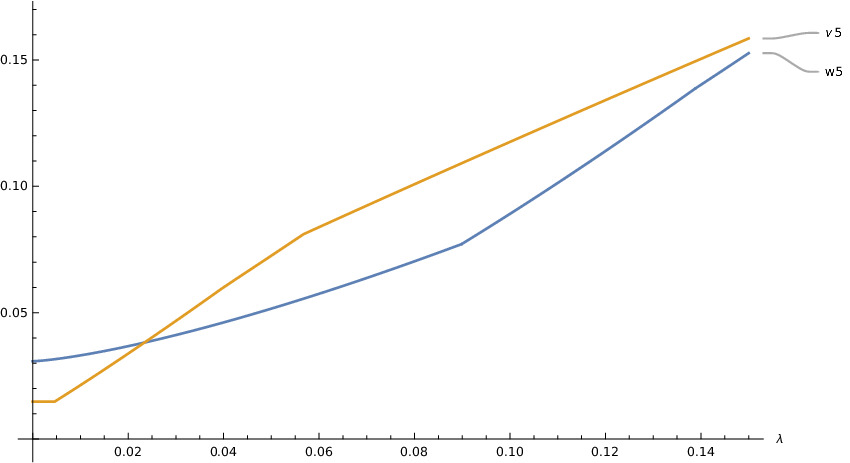}\hfill \includegraphics[scale=0.55]{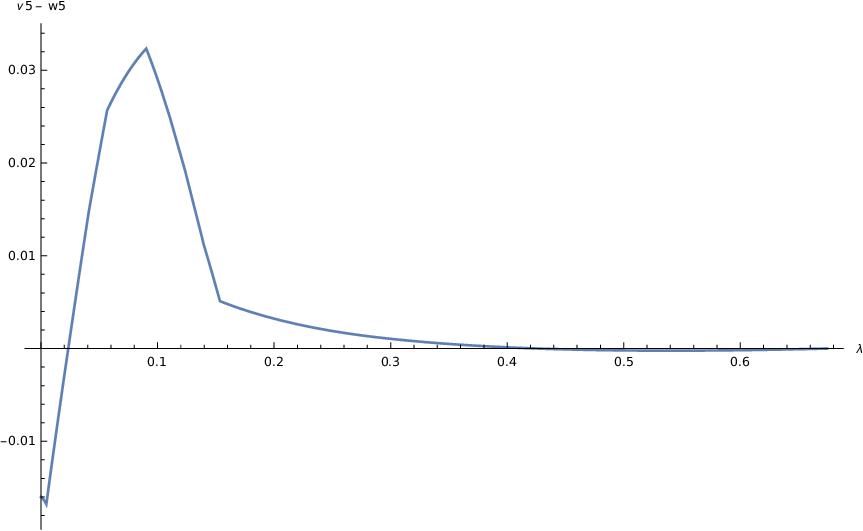}\hfill\null  
\caption{The nonperturbative flow of the running constants $\nu_5$ and $\ww_5$} \label{2507d}
\end{figure}
For Uranium atom we use only a part of the spectrum. This part is given in appendix~\ref{0611b} and corresponds to the occupied $s$-orbitals.
The remainder can be estimated as follows
\begin{align}
\sum \limits^{\infty}_{n>7} \frac{1}{Z}\int \limits^{x_n}_{x_{n+1}}dx  \, \ww^{\prime}_5=\frac{1}{Z} \int \limits^{\frac{1}{8}}_{0}dx  \, \ww^{\prime}_5=\frac{\ww_5(\frac{1}{8})-\ww_5(0)}{Z} = \frac{0.043}{Z}=4.7 \times 10^{-4}. \label{2707a}
\end{align}
Summing over first six spectral intervals in equation~\eqref{2507c} we obtain $\nu^{0 \Lambda}_5=0.015$. In the left panel of Figure~\ref{2507d}  we give the flows of the running constant corresponding to an atom~$\nu^{\lambda \Lambda}_5$ and a one-electron ion~$\ww^{\lambda \Lambda}_5$ of Uranium. The right panel contains the difference between two running constants $\nu^{\lambda \Lambda}_5 - \ww^{\lambda \Lambda}_5$. Finally, the density~\eqref{2507f} has the form
\begin{equation}
\nu^{0 \Lambda}(r)=\frac{1}{|r|^7}\left\{ \begin{matrix}6 \times 10^{-4} & \mbox{92 electrons},\\1.2 \times 10^{-3} &\mbox{one electron},\end{matrix}\right. \quad |r|>1. \label{0610a}
\end{equation}
Here $r$ is in the units of Compton length. For a one-electron ion with $Z=92$ the exact result is known. The corresponding density at one Compton length is $1.3 \times 10^{-3}$~\cite{wk}. However, as we see in~\eqref{0610a}, for an atom of Uranium the nonperturbative correction is one half of it.

\section{Acknowledgments}
The work has been realized for the project \href{https://anr.fr/Project-ANR-17-CE29-0004}{ANR-17-CE29-0004} founded by the French National Research Agency. We are grateful to T. Leininger for providing us with an access to a dedicated computational node and T. Saue for calculating the spectrum of Uranium in appendix~\ref{0611b}.

\begin{appendices}
\section{Functional integral}
\subsection{Integration by parts}
Let $f,\bar{f} \in \mathcal{G}(\mathbb{R}^4)$ Grassmann smooth functions with compact support; $\psi,\bar{\psi} \in \mathcal{G}^\prime(\mathbb{R}^4)$; and denote by $S$ the covariance of the measure~$d \mu$, see e.g.~\eqref{2002a},~\eqref{0503a}.
\begin{align}
  \int d \mu  \,  \langle \bar{f} \psi \rangle e^{ \langle \bar{\eta} \psi\rangle + \langle \bar{\psi} \eta \rangle}&= \langle \bar{f} \frac{\delta}{\delta \bar{s}} \rangle \int d \mu  \, e^{ \langle \bar{\eta} + \bar{s}, \psi\rangle + \langle \bar{\psi} \eta \rangle}\Big|_{\bar{s}=0} =\langle \bar{f} \frac{\delta}{\delta \bar{s}} \rangle e^{\langle \bar{\eta} + \bar{s},S \eta \rangle}\Big|_{\bar{s}=0} \nonumber \\
  &=\langle \bar{f} S \eta \rangle e^{\langle \bar{\eta}S \eta \rangle}= \int d \mu \, \langle \bar{f} S  \frac{\delta}{\delta \bar{\psi} }\rangle e^{ \langle \bar{\eta} \psi\rangle + \langle \bar{\psi} \eta \rangle},\\
  \int d \mu  \,  \langle \bar{\psi} f \rangle e^{ \langle \bar{\eta} \psi\rangle + \langle \bar{\psi} \eta \rangle}&= -\langle \frac{\delta}{\delta s} f \rangle \int d \mu  \, e^{ \langle \bar{\eta} \psi\rangle + \langle \bar{\psi}, \eta + s \rangle}\Big|_{s=0} =-\langle \frac{\delta}{\delta s} f \rangle e^{\langle \bar{\eta}S, \eta + s \rangle}\Big|_{s=0} \nonumber \\
&=\langle \bar{\eta} S f \rangle e^{\langle \bar{\eta}S \eta \rangle}= - \int d \mu \, \langle \frac{\delta}{\delta \psi } S f\rangle e^{ \langle \bar{\eta} \psi\rangle + \langle \bar{\psi} \eta \rangle}.
\end{align}
 It follows that for Grassmann valued functionals $F_{\psi \bar{\psi}} \in \mathrm{span}\{e^{ \langle \bar{\eta} \psi\rangle + \langle \bar{\psi} \eta \rangle }: f,\bar{f} \in \mathcal{G}\}$ we get
\begin{align}
  \langle \bar{f} \psi \rangle F_{\psi \bar{\psi}} &=  \langle \bar{f} S \frac{\delta}{\delta \bar{\psi}} \rangle F_{\psi \bar{\psi}}\;,& \langle \bar{\psi} f \rangle F_{\psi \bar{\psi}} &= - \langle \frac{\delta}{\delta \psi}  S f \rangle F_{\psi \bar{\psi}}\;. \label{0603a}
\end{align}
A similar formula holds for the Gaussian measure~\eqref{2911a}. For $F_A \in \mathrm{span}\{e^{ A(f)}: f \in \mathcal{D}\}$ one obtains
\begin{equation}
  \langle j A \rangle F_A = \langle j C \frac{\delta}{\delta A} \rangle F_A. \label{0803c}
\end{equation}
For details see~\cite{gljf}.
\subsection{Scaling properties} \label{2503a}
Let $d \mu_C$ denote a Gaussian measure with covariance $C$, see~\eqref{2911a},
\begin{align}
  Z(j)&=  \int d \mu_C \, e^{- \frac{1}{2}(Z_3 -1) \langle A C^{-1} A\rangle + \langle j A\rangle},& Z^\prime_j(j)&=\frac{\delta Z(j)}{\delta j}.
\end{align}
Using integration by parts~\eqref{0803c} we obtain
\begin{align}
  Z^\prime_j(j)&=\int d \mu_C \, (-(Z_3 -1) A + C j) e^{- \frac{1}{2}(Z_3 -1) \langle A C^{-1} A\rangle + \langle j A\rangle}\nonumber \\
  &=-(Z_3 -1) Z^\prime_j(j) + C j Z(j).
\end{align}
It follows that $Z(j)$ is the characteristic function of the measure $d \mu_{Z^{-1}_3C}$
\begin{equation}
  Z^\prime_j(j)= Z^{-1}_3 C j Z(j) \implies Z(j)=Z(0)e^{\frac{1}{2} \langle j Z^{-1}_3 C j\rangle}.
\end{equation}
Similarly let $d \mu_S$ denote a Grassmann measure with covariance~$S$, see~\eqref{2002a},
\begin{align}
  Z(\bar{\eta}, \eta)&=  \int d \mu_S \, e^{- (Z_2 -1) \langle \bar{\psi} S^{-1} \psi\rangle + \langle \bar{\psi} \eta\rangle + \langle \bar{\eta} \psi \rangle},& Z^\prime_{\bar{\eta}}(\bar{\eta},\eta)&=\frac{\delta Z(\bar{\eta},\eta)}{\delta \bar{\eta}}.
\end{align}
Using integration by parts~\eqref{0603a} we have
\begin{align}
  Z^\prime_{\bar{\eta}}(\bar{\eta}, \eta)&=\int d \mu_S \, (-(Z_2 -1) \psi + S \eta) e^{- (Z_2 -1) \langle \bar{\psi} S^{-1} \psi\rangle + \langle \bar{\psi} \eta\rangle + \langle \bar{\eta} \psi \rangle}\nonumber \\
                                         &=-(Z_2 -1) Z^\prime_{\bar{\eta}}(\bar{\eta}, \eta) + S \eta Z(\bar{\eta}, \eta),\\
  Z^\prime_{\eta}(\bar{\eta}, \eta)&=\int d \mu_S \, ((Z_2 -1) \bar{\psi} - \bar{\eta} S) e^{- (Z_2 -1) \langle \bar{\psi} S^{-1} \psi\rangle + \langle \bar{\psi} \eta\rangle + \langle \bar{\eta} \psi \rangle}\nonumber \\
                                         &=-(Z_2 -1) Z^\prime_{\eta}(\bar{\eta}, \eta) - \bar{\eta} S Z(\bar{\eta}, \eta).
\end{align}
This yields
\begin{align}
  Z^\prime_{\bar{\eta}}(\bar{\eta}, \eta)&= Z^{-1}_2 S \eta Z(\bar{\eta}, \eta),&  Z^\prime_{\eta}(\bar{\eta}, \eta)&= - \bar{\eta} Z^{-1}_2 S Z(\bar{\eta}, \eta).
\end{align}
Consequently $Z(\bar{\eta}, \eta)$ is the characteristic function of the measure $d \mu_{Z^{-1}_2 S}$
\begin{equation}
  Z(\bar{\eta}, \eta)=Z(0) e^{\langle \bar{\eta} Z^{-1}_2S \eta \rangle}.
\end{equation}
\section{The Laplace transform}
The Laplace transform restricted to an interval $[a,b]$ is the following map
\begin{equation}
f(x) \mapsto \hat{f}(p)=\int \limits^b_a dx \, f(x) e^{-px}.\label{2207b}
\end{equation}
We use the following transformations:
\begin{align}
e^{iwx} &\mapsto \hat{\sss}_\omega(p)=\frac{e^{-(p- iw)} - e^{-(p-iw)}}{p-iw}\,,\\
\frac{1}{x^{m+1}}& \mapsto \hat{\ee}_m(p)= a^{-m} E_{m+1}(ap)-b^{-m}E_{m+1}(p b)\,,\\
x&\mapsto \left(\frac{e^{-b p} - e^{- ap}}{p}\right)^\prime_p,\\
\uu(x) &\mapsto \hat{\uu}(p)= - \frac{8 \gamma}{3 \pi} \int \limits^\infty_1 d\zeta \,\sqrt{\zeta^2 -1}\left(1 + \frac{1}{2 \zeta}\right)\left( \frac{e^{-(2 \zeta + p)b} - e^{-(2 \zeta + p) a}}{p + 2 \zeta} \right)^\prime_\zeta\,,
\end{align}
where $E_n$ is the exponential integral function, the definition of~$\uu$ see in~\eqref{2207a}.

\section{Solutions of the radial Dirac equation}\label{1907d}
For $(\kappa,n) \in (\mathbb{Z}_+, \mathbb{N}_+) \cup (\mathbb{Z}_-, \mathbb{N})$ the eigenfunctions of the radial Dirac equation~\eqref{2405a} are
\begin{align}
\tilde{F}^{\kappa n}(r)&=(1+z)^{\frac{1}{2}} (2 p r)^s u_1 \sqrt{p} e^{-pr} \tilde{H}_-(r)\,,\\
\tilde{G}^{\kappa n}(r)&=-(1-z)^{\frac{1}{2}} (2 pr )^s u_1 \sqrt{p} e^{-pr}  \tilde{H}_+(r)\,,\\
\tilde{H}_\pm&=\left(\frac{n+s}{z} - \kappa\right)F_1(-n,2 s +1 ,2 pr) \pm n F_1(1-n,2s + 1 , 2pr)\,,
\end{align}
\begin{align}
p&=(1-z^2)^{\frac{1}{2}},& z&=\left(1 + \frac{\gamma^2}{(n+s)^2}\right)^{-\frac{1}{2}},\label{2107b}\\
s&=(\kappa^2 - \gamma^2)^{\frac{1}{2}},&u_1&=\frac{1}{\Gamma(2s +1)} \left(\frac{\Gamma(2s +n +1)}{2 \frac{n+s}{z} (\frac{n+s}{z} - \kappa) n!}\right)^{\frac{1}{2}}\,.
\end{align}     
For $\kappa \in \mathbb{Z}\backslash\{0\}$, $p >0$ solutions of the radial equation in~$L_\infty$ are
\begin{align}
F^{\kappa p}_\pm(r)&=\left(\frac{z+1}{z}\right)^{\frac{1}{2}}  (2pr)^s u_2\mathrm{Re} H(2i pr )\,,\\
G^{\kappa p}_\pm(r)&=\mp \left(\frac{z-1}{z}\right)^{\frac{1}{2}}  (2pr)^s  u_2 \mathrm{Im} H(2i pr )\,,\\
H(x)&=e^{-\frac{x}{2}} \left( \left(- \kappa + i \frac{y}{z} \right)(s + iy)\right)^{\frac{1}{2}} F_1(s+1 + iy,2s+1,x)\,,
\end{align}
\begin{align}
u_2&=\frac{e^{\frac{\pi y}{2}}|\Gamma(s +i y)|}{\sqrt{\pi} \Gamma(1+2s)},&y&=\frac{\gamma z}{p},& z&=\pm(1 + p^2)^{\frac{1}{2}},&s&=(\kappa^2 - \gamma^2)^{\frac{1}{2}}.
\end{align}
For details of calculations see~\cite{swain1,rose,greiner}.
\section{Spectrum $\sigma(\h)$ of Uranium} \label{0611b}
\begin{center}
\begin{tabular}{c|c}
  $n$& $p_n=\sqrt{1-z^2_n}$\\\hline
1&0.67466\\
2&0.29191\\
3&0.14666\\
4&0.07417\\
5&0.03530\\
6&0.01455\\
7&0.00458\\
\end{tabular}
\end{center}
\section{The vacuum polarization at short distances}
In Minkowski space the vacuum polarization at 1-loop is given by the following expression
\begin{align} 
  \hat{\Pi}^{\mu \nu}(p) &= i \int \limits_\Omega \frac{d^4 k}{(2 \pi)^4} \frac{Tr \{(- i \slashed{k} + m) \gamma^\mu (-i (\slashed{k} + \slashed{p})+ m) \gamma^\nu\}}{(k^2 + m^2 - i \epsilon)((k+p)^2 + m^2 -i \epsilon)},& \Omega&=\{ k \in \mathbb{R}^4: k^2 < \Lambda^2_0\}.
\end{align} 
One usually performs the Wick rotation to Euclidean space, calculates the traces and uses the Feynman parameterization
\begin{equation}
  \frac{1}{a_1 \dots a_n}=(n-1)!\int \limits^1_0 d x_1 \dots d x_n \frac{\delta(1- \sum^n_{i=1} x_i)}{(\sum^n_{i=1} a_i x_i)^n}.
\end{equation}
These steps yield the following form
\begin{equation}
\hat{\Pi}^{\mu \nu}(p)= 4i^2 \int \limits^1_0 dx \, (-2 I^{\mu \nu}_{x} - (p^\nu I^\mu_{x}  + p^\mu I^\nu_{x} ) + \delta^{\mu \nu} (I^2_{x} + p_\alpha I^\alpha_{x} + m^2 I^0_{x}))\label{0907a}
\end{equation}
where
\begin{align}
I^{\mu \nu}_{x} &= \int \limits_\Omega \frac{d^4 k}{(2 \pi)^4} \frac{k^\mu k^\nu}{(k^2 + 2 k p_x + M^2_x)^2}\,,&p_x&=p(1-x),&M^2_x&=p^2(1-x) + m^2.
\end{align}
The calculation of all integrals in~\eqref{0907a} is lengthy but straightforward. Eventually we obtain the following result
\begin{align}
\pi(p)&=\frac{1}{12 \pi^2}  \log \frac{\Lambda^2_0}{m^2} - \frac{1}{36 \pi^2} -\frac{1}{2 \pi^2} \int \limits^1_0 dx \, x(1-x) \log \frac{M^2_x}{m^2} \,,\label{2007a}\\
\hat{\Pi}^{\mu \nu}(p)&= (p^2 g^{\mu \nu} - p^\mu p^\nu)\pi(p) +  \frac{g^{\mu \nu}}{8 \pi^2} (m^2 - \Lambda^2_0) + \frac{p^\mu p^\nu}{24 \pi^2}\,.\label{0907b}
\end{align}
\section{Weyl basis}
\begin{align}
  \gamma^0 &=-i \begin{pmatrix}0&1\\1&0\end{pmatrix},&\gamma^i&=-i \begin{pmatrix} 0& \sigma_i \\-\sigma_i &0\end{pmatrix},&\gamma^5&=-i\gamma^0 \gamma^1 \gamma^2 \gamma^3=\begin{pmatrix}1&0\\0&-1\end{pmatrix}.\label{1204b}
\end{align}
\end{appendices}

\end{document}